\documentclass[tighten,twocolappendix,twocolumn]{aastex631}

\usepackage{multirow}
\usepackage{upgreek}
\usepackage{mathtools}
\usepackage{enumitem}

\usepackage{booktabs} 

\usepackage{hyperref}
\newcommand{\uat}[2]{\href{http://astrothesaurus.org/uat/#2}{#1 (#2)}}
\usepackage{rotating}

\begin{document}

\title{Refitting cosmological data with neutrino mass and degeneracy}

\author{Shek Yeung}
\author[0000-0003-0102-1543]{Wangzheng Zhang}
\author[0000-0002-1971-0403]{Ming-chung Chu}
\affiliation{Department of Physics, The Chinese University of Hong Kong, Sha Tin, NT, Hong Kong}

\email{terryys@gmail.com}
\email{wzzhang@link.cuhk.edu.hk}
\email{mcchu@phy.cuhk.edu.hk}

\begin{abstract}
A simple and natural extension of the standard Lambda cold dark matter ($\Lambda$CDM) model is to allow relic neutrinos to have finite chemical potentials. We confront this $\Lambda$CDM$\xi$ model, a $\Lambda$CDM with neutrino mass $M_\nu$ and degeneracy $\xi_3$ as additional parameters, with various cosmological data sets. We find that the $H_0$ and $S_8$ tensions become significant only in the presence of the cosmic microwave background (CMB) polarization data. Specifically, the global and local measurements agree to within 0.8$\sigma$ and 1.6$\sigma$ for the $H_0$ and $S_8$ tensions, respectively, when the CMB polarization data are not included. Therefore, the $H_0$ and $S_8$ tensions exist between CMB temperature and polarization data, both being global measurements. Fitting the $\Lambda$CDM$\xi$ model to the CMB temperature data, we find 3$\sigma$ evidence for nonzero neutrino mass ($M_\nu=0.57^{+0.17}_{-0.13}\,\mathrm{eV}$) and degeneracy ($\xi_3=1.13^{+0.41}_{-0.19}$), and the $\mathcal{O}(1)$ neutrino degeneracy parameter is compatible with Big Bang nucleosynthesis data. The scalar index $n_s$ exceeds 1 slightly, which is compatible with some hybrid inflation models. Furthermore, the recent DESI baryon acoustic oscillation data prefer the $\Lambda$CDM$\xi$ model to the Planck $\Lambda$CDM model. Similar results are obtained when including additional supernova data, while the inclusion of the Atacama Cosmology Telescope $\tau$ prior shifts the preferred $M_\nu$ and $\xi_3$ values closer to zero and brings $n_s$ back to the values favored when the polarization data are included.
\end{abstract}


\section{Introduction} \label{sc:intro}

The recent availability of high-precision cosmological and astrophysical data allows for testing cosmological models in unprecedented detail. The Lambda cold dark matter ($\Lambda$CDM) model has been very successful in accounting for a wide range of observations \citep{planck2020aa, ACT2020_s1, ACT2023_s2, SDSS2020, SPT2021, SPT2023, DES2022, SH0ES2022ApJ, asgari2021aa, brout2022apj}, though there are discordances among different fitting results. The most notable inconsistency is the Hubble tension, a significant discrepancy between the global and local measurements of the Hubble constant $H_0$. Specifically, there is about 5$\sigma$ disagreement between the result obtained by the Planck collaboration \citep{planck2020aa}, $H_0=67.35^{+0.54}_{-0.53}\,\mathrm{km}\,\mathrm{s}^{-1}\mathrm{Mpc}^{-1}$ at a 68\% confidence level (CL) derived from the anisotropies of the cosmic microwave background (CMB) TT, TE, EE, lowT, lowE and Plik lensing data, and that of the SH0ES collaboration \citep{SH0ES2022ApJ}, $H_0=73.04\pm1.04\,\mathrm{km}\,\mathrm{s}^{-1}\mathrm{Mpc}^{-1}$ at a 68\% CL measured using Type Ia supernovae calibrated by Cepheids.

There is another milder tension on the strength of matter clustering, quantified by the parameter $S_8 \equiv \sigma_8 \sqrt{\Omega_m/0.3}$, between the measurements through CMB and weak gravitational lensing and galaxy clustering. Here, $\Omega_m$ represents the matter density and $\sigma_8$ is the standard deviation of matter density fluctuations at radius $8\,h^{-1}\mathrm{Mpc}$ extrapolated to redshift 0 according to linear theory, where $h$ is the dimensionless Hubble parameter. Specifically, there is an approximate 3$\sigma$ discrepancy between $S_8=0.832\pm0.013$ at 68\% CL from Planck CMB using TT, TE, EE, lowT, lowE and Plik lensing data \citep{planck2020aa} and $S_8=0.759^{+0.024}_{-0.021}$ from KiDS-1000 weak lensing data \citep{asgari2021aa}. Moreover, weak lensing observations from the Hyper Suprime-Cam Subaru Strategic Program (HSC-SSP) indicate a slightly lower tension, ranging from $2\sigma$ to $2.5\sigma$ depending on the specific methodologies used \citep{HSC-SSP2023PhRvD_s1,HSC-SSP2023PhRvD_s2,HSC-SSP2023PhRvD_s3,HSC-SSP2023PhRvD_s4,HSC-SSP2023PhRvD_s5}.

These tensions could suggest the presence of unknown systematic errors in the cosmological data and/or new physics beyond $\Lambda$CDM. For the latter, a variety of new physics models have been proposed to alleviate both tensions together, such as interacting dark energy \citep{IDE2020_s1, IDE2020_s2, IDE2019_s5, IDE2021_s3}, decaying dark matter \citep{DDM2015_s1}, running vacuum model \citep{RVM2021_s1}, graduated dark energy \citep{gDE2021_s1, gDE2023_s2}, and $f(T)$ gravity \citep{fT2020_s1}, among others. More detailed discussions on the subject can be found in \cite{H0S8solution2022JHEAp_review, H0solution2021CQGra_review1, H0solution2022PhR_review2, khalife2023, H0review2024}. Nevertheless, resolving both tensions with the same new physics while preserving the successes of $\Lambda$CDM remains challenging.

In exploring these tensions, an important uncertainty lies in the value of the optical depth ($\tau$), which is primarily determined by the CMB polarization data. Recent observations from the James Webb Space Telescope (JWST) suggest a higher value of $\tau \gtrsim 0.07$ \citep{Julian2024MNRAS}, compared to the Planck value of $\tau = 0.0540\pm0.0074$ \citep{planck2020aa}, driven by the greater ionizing efficiency in early galaxies \citep{Simmonds2024MNRAS, Endsley2024MNRAS}. Additionally, an independent measurement by the Atacama Cosmology Telescope (ACT), which relies solely on ACT temperature, polarization, and lensing data combined with baryon acoustic oscillation (BAO) and supernova data, yields $\tau = 0.076 \pm 0.015$—a result that is independent of Planck and more consistent with the JWST estimate \citep{tau_wo_CMB2023}.

This uncertainty in $\tau$ provides room for new possibilities in resolving these tensions. Recent studies, such as \cite{allali2025reionization}, have shown that excluding large-scale polarization data can alleviate the $H_0$ tension. In this context, we focus specifically on the role of neutrinos in addressing the $H_0$ and $S_8$ tensions by considering the $\Lambda$CDM + $M_\nu$ + $\xi_i$ model ($\Lambda$CDM$\xi$), a minimal extension of the $\Lambda$CDM model, with $M_\nu\equiv\sum_im_i$ and $\xi_i\equiv\mu_i/T_\nu$ being the neutrino mass and degeneracy parameter, respectively, where $m_i$ and $\mu_i$ denote the mass eigenvalue and chemical potential of the $i$th neutrino mass eigenstate, respectively, and $T_\nu$ is the relic neutrino temperature. Hereafter, following the convention in \cite{planck2020aa}, $\Lambda$CDM represents the standard cosmological model with $M_\nu$ set to $0.06\,\mathrm{eV}$ and $\xi_i=0$. While the electron-type neutrino degeneracy parameter $\xi_e\ll1$ is tightly constrained by Big Bang nucleosynthesis \citep[BBN;][]{xinu_constrain2023_s1, xinu_constrain2023_s2}, it is still possible for $\mathcal{O}(1)$ degeneracy parameters for muon-type and tau-type neutrinos ($\nu_\mu, \nu_\tau$). Previous studies have shown that nonzero $\xi_i$ can alleviate the $H_0$ tension \citep{xinu2017EPJC, Yeung_2021, Osamu2021}. In this Letter, we revisit the impact of the $\Lambda$CDM$\xi$ model on the $H_0$ and $S_8$ tensions by comparing fits with and without CMB polarization data.

This Letter is structured as follows. First, we briefly review the neutrino properties and observational data that we used in this Letter in Sections \ref{sc:method} and \ref{sc:data}, respectively. Then, we summarize our main results related to the $H_0$ and $S_8$ tensions in Section \ref{sc:cmb_tensions}. Following this, Section \ref{sc:lcdmxi_TT} delves into the potential impacts of the $\Lambda$CDM$\xi$ model. Lastly, our conclusions are presented in Section \ref{sc:conclusion}.
\section{Methodology}\label{sc:method}

As shown in \cite{xinu2017EPJC}, the neutrino degeneracy becomes diagonal in the mass eigenbasis just before neutrino decoupling, and $\xi_i$ can be related to $\xi_\alpha$ in the flavor basis by the Pontecorvo–Maki–Nakagawa–Sakata matrix \citep[e.g., Eq.~(13) in ][]{xinu2017EPJC}, where $\alpha=e,\mu,\tau$. Based on strong constraints from BBN \citep{xinu_constrain2023_s1,xinu_constrain2023_s2} and the strong mixing between $\nu_\mu$ and $\nu_\tau$, we assume $\xi_e=0$ and $\xi_\mu=\xi_\tau$. Then, we are left with only one free parameter related to neutrino degeneracy, which we choose to be $\xi_3$.

The neutrino degeneracy parameters enter the CMB anisotropy calculation through the neutrino energy density, 
\begin{equation}\label{eq:rho_neu}
    \rho_\nu = \frac{1}{2\uppi^2}\sum_{i=1}^3
    \int_0^\infty
    \left[
    \frac{\sqrt{p^2+m_i^2}}{e^{p/T_\nu-\xi_i}+1} + \frac{\sqrt{p^2+m_i^2}}{e^{p/T_\nu+\xi_i}+1}
    \right]
    p^2dp,
\end{equation}
which plays a role in the expansion history of the Universe and the Boltzmann equations. Following \cite{Yeung_2021}, we modified CAMB \citep{CAMB1,CAMB2} to include these effects in the calculations of the CMB anisotropy power spectra. More details can be found in \cite{Yeung_2021}. In this Letter, we consider the CMB fittings of the following two models: $\Lambda$CDM, with $M_\nu=0.06\,\mathrm{eV}$ and $\xi_3=0$; and $\Lambda$CDM$\xi$, in which $M_\nu$ and $\xi_3$ are allowed to vary.

\section{Observational Data Sets} \label{sc:data}

The observational data sets used in the Markov Chain Monte Carlo (MCMC) analyses in this Letter are as follows:
\begin{enumerate}[noitemsep,topsep=0pt]
    \item \textit{CMB temperature and lensing.} The high-multipole ($\ell > 30$) CMB temperature anisotropy power spectra (TT) from the \texttt{plik} Planck 2018 likelihood, the low-multipole ($2 \leq \ell \leq 30$) temperature power spectrum (lowT) from \texttt{Commander}, and the CMB lensing reconstruction (Plik lensing) \citep{planckV2020A&A,planckVIII2020A&A}.
    \item \textit{CMB polarization.} The high-multipole ($\ell > 30$) CMB polarization anisotropy power spectra (TE, EE) from \texttt{plik} and the low-multipole ($2 \leq \ell \leq 30$) polarization power spectrum (lowE) from \texttt{SimAll}\citep{planckV2020A&A}.
    \item \textit{BAO.} BAO data used in the Planck 2018 analysis \citep{planck2020aa}, including measurements from BOSS DR12 \citep[LOWZ and CMASS,][]{bao2017mnras}, the 6dF Galaxy Survey \citep{bao2011mnras}, and the Sloan Digital Sky Survey DR7 Main Galaxy Sample \citep{bao2015mnras}. A full list is provided in Appendix~\ref{sc:semi_analysis_BAO}, Table~\ref{tab:BAO_data}.
    \item \textit{Weak lensing.} The first year data from the Dark Energy Survey \citep[DES Y1,][]{des2018prd} used in the Planck 2018 analysis \citep{planck2020aa}.
    \item \textit{Supernova.} The Dark Energy Survey Supernova 5YR data \citep{Abbott_2024, Sanchez_2024, Vincenzi_2024}.
\end{enumerate}

The observational data used to estimate cosmological tensions include:
\begin{enumerate}[noitemsep,topsep=0pt]
    \setcounter{enumi}{5}  
    \item \textit{SH0ES.} The recent $H_0$ measurement from the SH0ES collaboration, $H_0 = 73.04 \pm 1.04$ km s$^{-1}$ Mpc$^{-1}$ \citep{brout2022apj}.
    \item \textit{KiDS.} Since $S_8$ is model dependent, we reanalyze the KiDS-1000 real-space shear two-point correlation functions \citep[2PCFs;][]{asgari2021aa}, obtaining $S_8 = 0.759^{+0.021}_{-0.025}$ for the $\Lambda$CDM model and $S_8 = 0.748^{+0.019}_{-0.022}$ for the $\Lambda$CDM$\xi$ model. See Appendix \ref{sc:S8_MCMC} for further details.
\end{enumerate}

For clarity, we define the following data set combinations used in the main text:

\begin{enumerate}[noitemsep,topsep=0pt]
    \item $\mathtt{P18}$: The standard Planck 2018 data set, including \textit{CMB temperature and lensing} and \textit{CMB polarization} data.
    \item $\mathtt{TTBW}$: \textit{CMB temperature and lensing} combined with \textit{BAO} and \textit{Weak lensing} data. Since $\tau$ is loosely constrained without polarization data, we consider additional $\tau$ priors. More details are in Section~\ref{sc:cmb_tensions}.
    \item $\mathtt{TEBW}$: \textit{CMB temperature and lensing}, \textit{CMB polarization}, \textit{BAO}, and \textit{Weak lensing} data.
    \item $\mathtt{TTBW+SN}$: $\mathtt{TTBW}$ and \textit{supernova} data.
\end{enumerate}

Hereafter, for simplicity, we refer to the fitting of model $\mathcal{M}$ to the data set $\mathcal{D}$ ($\mathtt{P18}$, $\mathtt{TTBW}$, or $\mathtt{TEBW}$) as $\mathcal{M}_\mathcal{D}$. We use CosmoMC \citep{COSMOMC1, COSMOMC2} to perform the MCMC analyses. The Gelman-Rubin statistic $R$ is used to determine the convergence of the MCMC chains. We consider $R-1<0.01$ as an indication of convergence.

\section{Tensions within CMB?}\label{sc:cmb_tensions}

\begin{table*}
    \renewcommand{\arraystretch}{1.3} 
    \setlength{\tabcolsep}{5pt}      
    \begin{tabular}{lcccccc}
    \hline\hline
         & \multicolumn{3}{c}{$H_0\,[\mathrm{km}\,\mathrm{s}^{-1}\mathrm{Mpc}^{-1}]$} 
         & \multicolumn{3}{c}{$S_8$} \\
         & $\mathtt{P18}$ & $\mathtt{TTBW}$ & $\mathtt{TEBW}$ 
         & $\mathtt{P18}$ & $\mathtt{TTBW}$ & $\mathtt{TEBW}$ \\
    \hline
    \multirow{2}{*}{$\Lambda\mathrm{CDM}$} 
         & $67.35^{+0.54}_{-0.53}$& $68.79\pm0.55$& $68.17\pm0.38$ 
                         & $0.832\pm0.013$ & $0.813\pm0.010$ & $0.8118^{+0.0090}_{-0.0092}$ \\
         & $4.9$          & $3.6$          & $4.4$ 
                         & $2.7$& $2.2$          & $2.1$\\
    \hline
    \multirow{2}{*}{$\Lambda\mathrm{CDM}\xi$} 
         & $67.5^{+1.2}_{-0.87}$            & $71.2\pm2.1$& $68.57^{+0.56}_{-0.62}$ 
                         & $0.831\pm0.013$            & $0.787\pm0.014$ & $0.8134^{+0.0098}_{-0.0097}$ \\
         & $3.7$            & $0.8$          & $3.5$ 
                         & $3.5$& $1.6$& $2.9$\\
    \hline
    \multirow{2}{*}{$\Lambda\mathrm{CDM}\xi+\tau_1$}& ---& $68.71^{+0.64}_{-0.89}$& ---& ---& $0.809\pm0.011$& ---\\
         & & $3.0$& & & $2.6$& \\
    \hline
    \multirow{2}{*}{$\Lambda\mathrm{CDM}\xi+\tau_2$}& ---& $68.75^{+0.82}_{-1.1}$& ---& ---& $0.806\pm0.012$& ---\\
         & & $2.8$& & & $2.4$& \\
    \hline\hline
    \end{tabular}
    \caption{Mean values (68\% CL) for $H_0$ (left) and $S_8$ (right) obtained from different data sets ($\mathtt{P18}$, $\mathtt{TTBW}$, and $\mathtt{TEBW}$) and models ($\Lambda\mathrm{CDM}$ and $\Lambda\mathrm{CDM}\xi$). Rows (3) and (4) are the results of adding two different $\tau$ priors to the $\Lambda\mathrm{CDM}\xi$ model, with $\tau_1 = \mathcal{N}(0.0540, 0.0074)$ and $\tau_2 = \mathcal{N}(0.076, 0.015)$. The numbers below the mean values are their deviations $n_\sigma$ (in units of $\sigma$) from those of the local measurements.}
    \label{tab:H0_S8_table}
\end{table*}

With $\Lambda\mathrm{CDM}_\mathtt{TTBW}$, we obtain (68\% CL): 
\begin{equation}
    \begin{aligned}
        H_0 &= 68.79\pm0.55\,\mathrm{km}\,\mathrm{s^{-1}}\,\mathrm{Mpc}^{-1}, \\
        S_8 &= 0.813\pm0.010.
    \end{aligned}
\end{equation}
To compare with local measurements, we calculate the deviations, in unit of standard deviations, $n_\sigma(H_0)=3.6$ and $n_\sigma(S_8)=2.2$, relative to the SH0ES measurement \citep[$H_0=73.04\pm1.04$ km s$^{-1}$ Mpc$^{-1}$,][]{brout2022apj} and KiDS results \citep[$S_8=0.759^{+0.024}_{-0.021}$,][]{asgari2021aa}, respectively, using the Marginalized Posterior Compatibility Level \citep{khalife2023}. Although these tensions are slightly reduced compared to those in $\Lambda\mathrm{CDM}_\mathtt{P18}$, with $n_\sigma(H_0)=4.9$ and $n_\sigma(S_8)=2.7$, they remain statistically significant.

\begin{figure*}
\centering
    \includegraphics[width=0.75\linewidth]{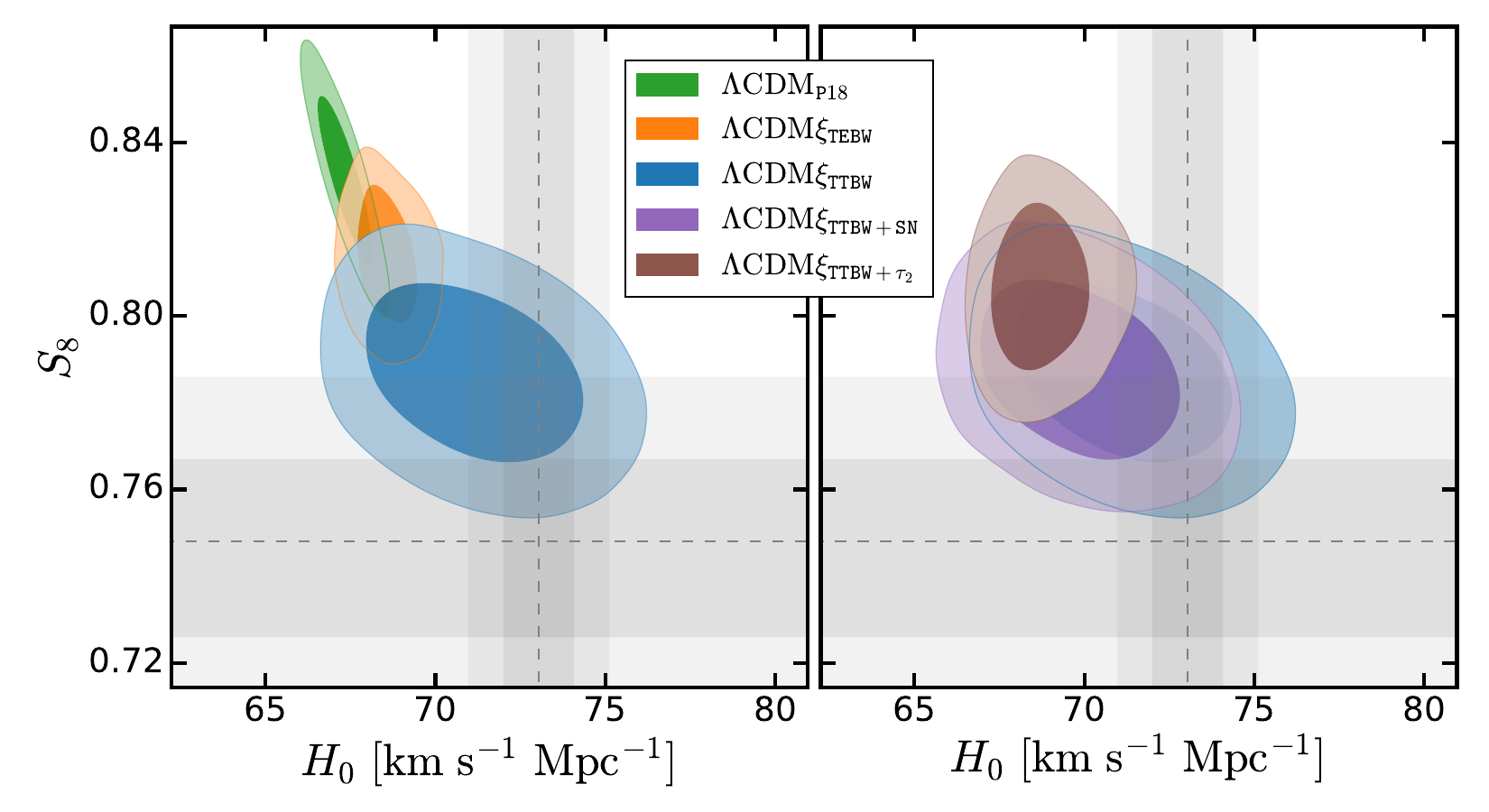}
    \caption{2D contours (for 68\% CL and 95\% CL) for $H_0$ and $S_8$ from $\Lambda\mathrm{CDM}_\mathtt{P18}$, $\Lambda\mathrm{CDM}\xi_\mathtt{TEBW}$, $\Lambda\mathrm{CDM}\xi_\mathtt{TTBW}$, $\Lambda\mathrm{CDM}\xi_{\mathtt{TTBW}+\mathtt{SN}}$ and $\Lambda\mathrm{CDM}\xi_\mathtt{TTBW+\tau_2}$. The dark and light gray regions represent the 68\% and 95\% CL, respectively, for the SH0ES $H_0$ measurement and the KiDS-1000 $S_8$ constraint under the $\Lambda$CDM$\xi$ model (see Section~\ref{sc:data}).}
    \label{fig:H0_S8_contour}
\end{figure*}

With $\Lambda$CDM$\xi_\mathtt{TTBW}$, we have (68\% CL)
\begin{equation}
    \begin{aligned}
        H_0 &= 71.2\pm2.1\,\mathrm{km}\,\mathrm{s^{-1}}\,\mathrm{Mpc}^{-1}, \\
        S_8 &= 0.787\pm0.014,
    \end{aligned}
\end{equation}
with $n_\sigma(H_0) = 0.8$ and $n_\sigma(S_8) = 1.6$. The $H_0$ and $S_8$ tensions are significantly alleviated when the neutrino mass $M_\nu$ and degeneracy $\xi_3$ are included as parameters. When the supernova data are included, with $\Lambda$CDM$\xi_\mathtt{TTBW+SN}$ we have (68\% CL)
\begin{equation}
    \begin{aligned}
        H_0 &= 69.9_{-2.1}^{+1.8}\,\mathrm{km}\,\mathrm{s^{-1}}\,\mathrm{Mpc}^{-1}, \\
        S_8 &= 0.787\pm0.014,
    \end{aligned}
\end{equation}
with $n_\sigma(H_0) = 1.5$ and $n_\sigma(S_8) = 1.6$, which is also shown in the right panel of Figure~\ref{fig:H0_S8_contour}. The inclusion of supernova data only slightly affects the results of $\Lambda$CDM$\xi_\mathtt{TTBW}$.

To compare the two models, we calculate the Bayes factor ($B$):
\begin{equation}\label{eq:bayes_factor}
    B = \frac{K({\Lambda\mathrm{CDM}}_\mathtt{TTBW})}{K({\Lambda\mathrm{CDM}\xi}_\mathtt{TTBW})},
\end{equation}
where $K$ is the Bayesian evidence calculated using \texttt{MCEvidence} \citep{heavens2017MCevidence}. We obtain $\log_{10}B=0.4$, indicating that the $\Lambda$CDM$\xi$ model fits the $\mathtt{TTBW}$ data almost as well as the $\Lambda$CDM model \citep{Bayes1995}.

However, when we include the CMB polarization data (i.e., $\mathtt{TEBW}$ data), the $H_0$ and $S_8$ tensions return, even with the $\Lambda$CDM$\xi$ model. Specifically, for $\Lambda\mathrm{CDM}_\mathtt{TEBW}$, we obtain $n_\sigma(H_0)=4.4$ and $n_\sigma(S_8)=2.1$, while for $\Lambda$CDM$\xi_\mathtt{TEBW}$, we have $n_\sigma(H_0) = 3.5$ and $n_\sigma(S_8) = 2.9$. More details are provided in Table~\ref{tab:H0_S8_table}. 

Without the CMB polarization data, the weak constraint on the optical depth $\tau$---due to degeneracies with the scalar amplitude $A_s$---allows the MCMC chains to explore a broader region of the parameter space. This results in higher values of $\tau$, such as $\tau=0.201^{+0.045}_{-0.037}$ for $\Lambda$CDM$\xi_\mathtt{TTBW}$ and $\tau=0.079\pm0.016$ for $\Lambda$CDM$_\mathtt{TTBW}$, leading to the reduction in $H_0$ and $S_8$ tensions. However, including the CMB polarization data tightens the constraints on $\tau$ and leads to lower values, such as $\tau=0.0542^{+0.0073}_{-0.0072}$ for $\Lambda$CDM$_\mathtt{P18}$, which reintroduces the tensions.

To further investigate this, we apply different priors for $\tau$ in $\Lambda\mathrm{CDM}\xi_\mathtt{TTBW}$: $\tau_1 = \mathcal{N}(0.0540, 0.0074)$ \citep[Planck \texttt{Plik} and \texttt{CamSpec} likelihood combined results,][]{planck2020aa} and $\tau_2 = \mathcal{N}(0.076, 0.015)$ \citep[ACT results, independent of Planck,][]{tau_wo_CMB2023}. With the $\tau_1$ prior, we find $H_0 = 68.71^{+0.64}_{-0.89}$ km s$^{-1}$ Mpc$^{-1}$ ($3.0\sigma$ tension) and $S_8 = 0.809 \pm 0.011$ ($2.6\sigma$ tension), along with a Bayes factor $\log_{10}B = -3.1$ compared to $\Lambda$CDM$\xi_\mathtt{TTBW}$ without $\tau$ prior. Using the $\tau_2$ prior, we obtain $H_0 = 68.75^{+0.82}_{-1.1}$ km s$^{-1}$ Mpc$^{-1}$ ($2.8\sigma$ tension) and $S_8 = 0.806 \pm 0.012$ ($2.4\sigma$ tension), with a Bayes factor $\log_{10}B = -2.0$. These results indicate that larger $\tau$ priors consistently alleviate both the $H_0$ and $S_8$ tensions, although neither prior fully resolves them. Additionally, the Bayes factor suggests that neither prior is favored by the $\mathtt{TTBW}$ data over the original $\Lambda\mathrm{CDM}\xi_\mathtt{TTBW}$. The results including the $\tau_2$ prior (hereafter referred to as $\Lambda$CDM$\xi_{\mathtt{TTBW}+\tau_2}$) are shown in the right panel of Figure~\ref{fig:H0_S8_contour}.

These results indicate that the $H_0$ and $S_8$ tensions arise not only from conflicts between local and global measurements but also from discrepancies within the CMB data itself, specifically between temperature and polarization data. This is clearly seen in the left panel of Figure~\ref{fig:H0_S8_contour}, where $\Lambda\mathrm{CDM}\xi_\mathtt{TTBW}$ is much closer to local measurements compared to $\Lambda\mathrm{CDM}\xi_\mathtt{TEBW}$. A summary of parameter constraints and 2D contours is provided in Table~\ref{tab:full_cosmo} and Figure~\ref{fig:full_CMB_contours}, respectively, in Appendix \ref{sc:cosmo_tables}.

\section{
Other implications of \texorpdfstring{$\Lambda$CDM$\xi_\mathtt{TTBW}$}{LCDMxiTTBW}
}\label{sc:lcdmxi_TT}
We systematically investigate the updated data sets, including the CMB temperature and lensing data from Planck 2020 \citep{PR42024}, the DESI BAO data \citep{desi2025arXiv250314745D, desi2025arXiv250314739D}, and the DES Y3 data \citep{DES2022}, individually, in Appendix \ref{sc:robustness}. Our analysis confirms that the results of the $\Lambda\mathrm{CDM}\xi_\mathtt{TTBW}$ model, which significantly alleviates the $H_0$ and $S_8$ tensions, remain robust across these data sets. Moreover, since the $\Lambda$CDM$\xi$ model fits the $\mathtt{TTBW}$ data as well as the $\Lambda$CDM model, but significantly reduces $H_0$ and $S_8$ tensions, we study the implications of $\Lambda$CDM$\xi_\mathtt{TTBW}$ on other cosmological parameters and the recent DESI BAO data \citep{desi2025arXiv250314745D, desi2025arXiv250314739D}.
\subsection{
Neutrino parameters: 
\texorpdfstring{$M_\nu$}{Mnu}
and 
\texorpdfstring{$\xi_3$}{xi3}
}\label{sc:Mnu_xi3_H0}

For $\Lambda$CDM$\xi_\mathtt{TTBW}$, we find that cosmological data prefers $M_\nu=0.57^{+0.17}_{-0.13}\,\mathrm{eV}$, $\xi_3=1.13^{+0.41}_{-0.19}$ (68\% CL), with a 3.3$\sigma$ (2.6$\sigma$) evidence for the nonzero $M_\nu$ ($\xi_3$). 

For $\Lambda$CDM$\xi_{\mathtt{TTBW}+\mathtt{SN}}$, we obtain $M_\nu = 0.61^{+0.15}_{-0.13}\,\mathrm{eV}$ and $\xi_3 = 1.05^{+0.42}_{-0.21}$ (68\% CL), corresponding to 4.0$\sigma$ and 2.1$\sigma$ evidence for nonzero $M_\nu$ and $\xi_3$, respectively. These results are broadly consistent with the $\Lambda$CDM$\xi_\mathtt{TTBW}$ case, indicating a strong preference for nonzero values of both neutrino parameters. In contrast, for $\Lambda$CDM$\xi_{\mathtt{TTBW}+\tau_2}$, we find $M_\nu = 0.167^{+0.083}_{-0.097}\,\mathrm{eV}$ and $\xi_3 = 0.40^{+0.12}_{-0.40}$ (68\% CL), with reduced significance levels of 1.8$\sigma$ and 0.5$\sigma$, respectively. This reduction is expected, as the tighter constraint on $\tau$ limits the freedom for neutrino parameters to vary.

\begin{figure*}
    \centering
    \includegraphics[width=0.75\linewidth]{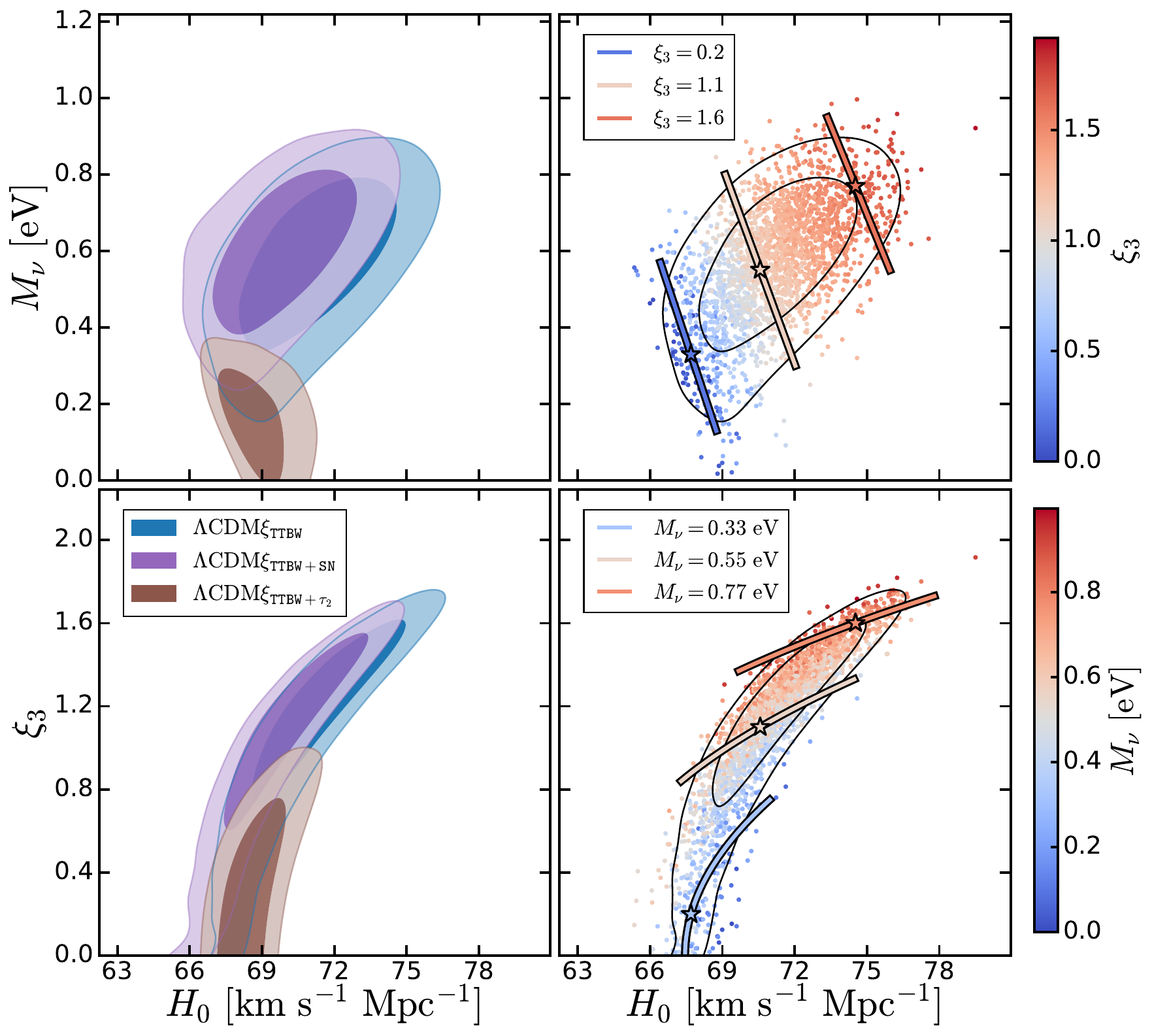}
    \caption{2D contours (68\% and 95\% CL) for $H_0$ and $M_\nu$ ($\xi_3$) from $\Lambda$CDM$\xi_\mathtt{TTBW}$, $\Lambda$CDM$\xi_{\mathtt{TTBW}+\mathtt{SN}}$, $\Lambda$CDM$\xi_{\mathtt{TTBW}+\tau_2}$ (left panels), and $\Lambda$CDM$\xi_\mathtt{TTBW}$ (right panels), with scatter points colored according to values of $\xi_3$ ($M_\nu$) in the upper (lower) panels. The colored lines are from Eqs. (\ref{eq:CH_Cm}) and (\ref{eq:CH_Cxi}), with the stars representing the corresponding reference points.}
    \label{fig:taylor_results}
\end{figure*}

The 2D contours (68\% and 95\% CL) for $H_0$ versus $M_\nu$ and $H_0$ versus $\xi_3$ for the above cases are shown in the left panels of Figure~\ref{fig:taylor_results}. To understand the correlations between $H_0$ and the neutrino parameters, we adopt a perturbative approach\footnote{Here, we use the $\Lambda$CDM$\xi_\mathtt{TTBW}$ case as an example; the same logic applies to the other cases.}---see Appendix \ref{sc:semi_analysis_theta} for details---based on the sound horizon angular size $\theta_*=r_s/D_M(z_*)$ at photon decoupling ($z_*$), which is constrained very well by the CMB data, where $r_s$ and $D_M(z_*)$ are the comoving sound horizon and distance to the last scattering, respectively. A small variation of $M_\nu$ around a reference value $M_\nu^\mathrm{ref}$ would induce corresponding variations in $r_s$ and $D_M$. We assume that $H_0$ is varied accordingly so as to keep $\theta_*$ unchanged (fixed to the observed value). We therefore obtain
\begin{equation}\label{eq:CH_Cm}
        C_H \frac{H_0 - H_0^\mathrm{ref}}{H_0^\mathrm{ref}} 
        = C_m \frac{M_\nu - M_\nu^\mathrm{ref}}{M_\nu^\mathrm{ref}},
\end{equation}
where $C_H$ and $C_m$ can be calculated using the expressions for $r_s$ and $D_M$. Similarly, we obtain the correlation between $H_0$ and $\xi_i$ by fixing $M_\nu$ together with $\theta_*$ and all other cosmological parameters:
\begin{equation}\label{eq:CH_Cxi}
        C_H \frac{H_0 - H_0^\mathrm{ref}}{H_0^\mathrm{ref}} 
        = 
        \sum_i C_{\xi,i}^1 (\xi_i - \xi_i^\mathrm{ref})
        + \sum_i C_{\xi,i}^2 (\xi_i - \xi_i^\mathrm{ref})^2.
\end{equation}
The coefficients $C_H, C_m, C_{\xi,i}^1, C_{\xi,i}^2$ are listed in Appendix \ref{sc:semi_analysis_theta}, Table~\ref{tab:taylor_calcoe}. These semianalytic results are shown in the right panels of Figure~\ref{fig:taylor_results}, which agree well with the MCMC results. Moreover, we can obtain roughly the same 95\% contours in the MCMC when the BAO data are included; see Appendix \ref{sc:semi_analysis_BAO} for details. The success of the perturbation theory indicates that the dominant reason for the correlations between the neutrino parameters and $H_0$ is due to their effects on the expansion history of the Universe.

Among the three cases considered, the $\Lambda$CDM$\xi_\mathtt{TTBW}$ and $\Lambda$CDM$\xi_{\mathtt{TTBW}+\mathtt{SN}}$ scenarios yield similarly strong indications of nonzero neutrino parameters, with $\xi_3 \sim 1$ and the significance above $2\sigma$. In contrast, the $\Lambda$CDM$\xi_{\mathtt{TTBW}+\tau_2}$ case shows only marginal evidence for $\xi_3$ (about $0.5\sigma$). Therefore, in the following BBN consistency check, we focus on the $\Lambda$CDM$\xi_\mathtt{TTBW}$ case as a representative example for scenarios with large $\xi_3$, where the implications for the early Universe are most relevant. From $\Lambda$CDM$\xi_\mathtt{TTBW}$, we find a larger $\Omega_bh^2$ ($0.02325^{+0.00042}_{-0.00039}$, 68\% CL), compared with $\Omega_bh^2=0.02242\pm0.00020$ from $\Lambda\mathrm{CDM}_\mathtt{TTBW}$. To estimate the implication of a larger $\Omega_bh^2$ on BBN, we adopt the following fitting formulae \citep{pitrou2018bbn} for the helium-4 mass fraction $Y_\mathrm{P}$ and deuterium-to-hydrogen ratio $\mathrm{D}/\mathrm{H}$:
\begin{equation}\label{eq:BBN_estimator}
    \begin{aligned}
        \frac{\Delta Y_\mathrm{P}}{\overline{Y_\mathrm{P}}} 
            &= 
                0.04\frac{\Delta \Omega_b h^2}{\overline{\Omega_b h^2}} 
              + 0.16\frac{\Delta N_\nu}{3.0} 
              - 0.96\xi_{\nu_e}, \\
        \frac{\Delta \mathrm{D}/\mathrm{H}}{\overline{\mathrm{D}/\mathrm{H}}} 
            &= 
              - 1.65\frac{\Delta \Omega_b h^2}{\overline{\Omega_b h^2}} 
              + 0.41\frac{\Delta N_\nu}{3.0} 
              - 0.53\xi_{\nu_e}.
    \end{aligned}
\end{equation}
The barred symbols represent the corresponding reference values used in \cite{pitrou2018bbn}. To keep the measured quantities of BBN ($Y_\mathrm{P}$ and $\mathrm{D}/\mathrm{H}$) unchanged, i.e., the left-hand sides of Eq.~(\ref{eq:BBN_estimator}) equal zero, a larger baryon density $\Omega_bh^2$ can be compensated by a larger effective number of neutrino species $N_{\nu}$ and the electron neutrino degeneracy parameter $\xi_{\nu_e}$. With $\Delta \Omega_b h^2 = 0.001$, we obtain $\Delta N_\nu=0.71$ and $\xi_{\nu_e}=0.04$, with the latter agreeing with recent fittings using new BBN data \citep[$\xi_{\nu_e}\sim0.04$,][]{xinu_constrain2023_s1,xinu_constrain2023_s2}. We also show in Appendix \ref{sc:BBN_He3} that with $\Delta N_\nu=0.71$ and $\xi_{\nu_e}=0.04$, the fractional deviation of the helium-3-to-hydrogen ratio $\frac{\Delta\prescript{3}{}{\mathrm{He}}/\mathrm{H}}{\overline{\prescript{3}{}{\mathrm{He}}/\mathrm{H}}}$ is negligible.

During the BBN era, neutrinos are relativistic particles, so we have 
\begin{equation}
    \Delta N_\nu = \frac{15}{7} \sum_i \left(\frac{\xi_i}{\uppi}\right)^2 \left[2+\left(\frac{\xi_i}{\uppi}\right)^2\right].
\end{equation}
We thus obtain $\xi_3=1.11$. Therefore, the mean values of $\xi_3$ and $\Omega_b h^2$ from $\Lambda$CDM$\xi_\mathtt{TTBW}$ are consistent with BBN data.

\subsection{
scalar index 
\texorpdfstring{$n_s$}{ns}\label{sc:ns_r}
}

\begin{figure*}
    \centering
    \includegraphics[width=0.75\linewidth]{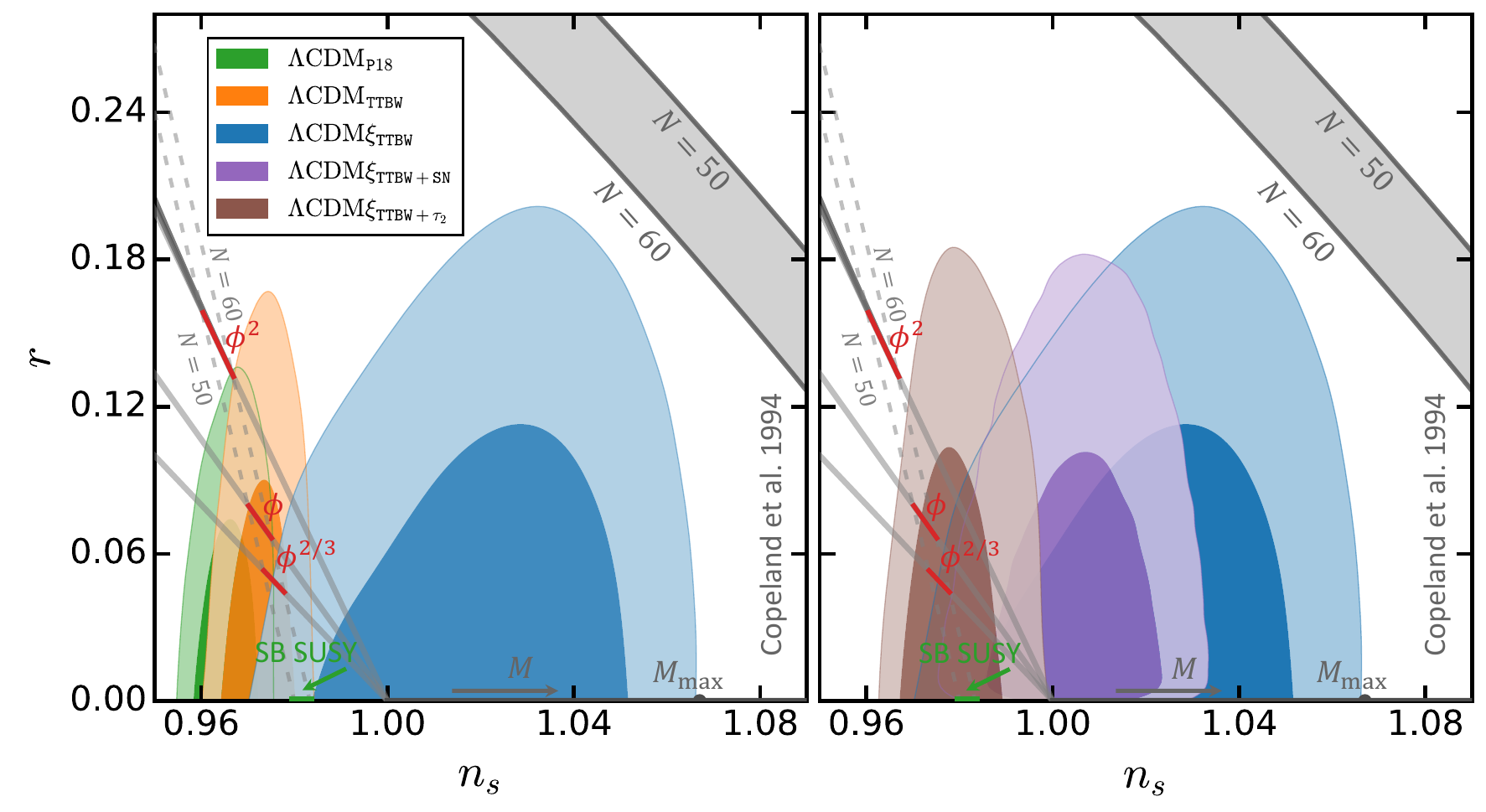}
    \caption{$n_s-r$ contours (for 68\% CL and 95\% CL) and several selected inflation models' predictions, shown as red, green, and dark gray lines, with uncertainties in the number of e-folds of $50<N<60$ (more details can be found in Section~\ref{sc:ns_r}). $M_\mathrm{max}$ is the upper bound of the model parameter $M$ in \cite{hybrid_inflation1994_s2}.}
    \label{fig:ns_r}
\end{figure*}

The mean value of $n_s$ in $\Lambda\mathrm{CDM}\xi_\mathtt{TTBW}$ ($n_s=1.019^{+0.022}_{-0.020}$ 68\% CL) is consistent with 1 and much higher than that of $\Lambda\mathrm{CDM}_\mathtt{TTBW}$ ($n_s=0.9721^{+0.0046}_{-0.0045}$ 68\% CL). Among the other extended cases, $\Lambda$CDM$\xi_{\mathtt{TTBW}+\mathtt{SN}}$ shows a similar behavior with $n_s = 1.013 \pm 0.020$ (68\% CL), while $\Lambda$CDM$\xi_{\mathtt{TTBW}+\tau_2}$ gives $n_s = 0.9777^{+0.0048}_{-0.0073}$ (68\% CL), closer to the standard Planck result.

To quantify the implications of our results on inflation models, we vary one more parameter, the scalar-to-tensor ratio $r$, when fitting the $\mathtt{TTBW}$ data. The 2D contours of $n_s-r$ for $\Lambda\mathrm{CDM}_\mathtt{P18}$, $\Lambda\mathrm{CDM}_\mathtt{TTBW}$, $\Lambda\mathrm{CDM}\xi_\mathtt{TTBW}$, $\Lambda$CDM$\xi_{\mathtt{TTBW}+\mathtt{SN}}$, and $\Lambda$CDM$\xi_{\mathtt{TTBW}+\tau_2}$ are shown in Figure~\ref{fig:ns_r}. The dark gray curve shows one of the simplest hybrid models of \cite{hybrid_inflation1994_s2}, with the number of e-folds $N$ between 50 and 60, where the energy density during inflation is dominated by the potential composed of two scalar fields $\phi$ and $\psi$,
\begin{equation}
    V(\phi,\psi)=\frac{1}{4}\lambda(\psi^2-M^2)^2 
                + \frac{1}{2}m^2\phi^2 
                + \frac{1}{2}\lambda'\phi^2\psi^2,
\end{equation}
where we take $\lambda=\lambda'=1$ following \cite{hybrid_inflation2009_s3} and fix the primordial perturbation amplitude $A_s$ according to our CMB fitting results \citep{hybrid_inflation1994_s2} so that we are left with only one parameter $M$ (or $m$). We obtain the upper bound for $M$ ($M_\mathrm{max}$) (or $m$ ($m_\mathrm{max}$)) around $1\times10^{-4}\,m_\mathrm{Pl}$ ($1\times10^{-8}\,m_\mathrm{Pl}$), where $m_\mathrm{Pl}\equiv\sqrt{\hbar c/G}$ is the Planck mass. Moreover, other hybrid models, such as those discussed in \cite{hybrid_inflation1994_s1, hybrid_inflation1994_s2, hybrid_inflation2009_s3}, can also be compatible with the results of $\Lambda\mathrm{CDM}\xi_\mathtt{TTBW}$ and $\Lambda\mathrm{CDM}\xi_{\mathtt{TTBW}+\mathtt{SN}}$. Furthermore, another kind of hybrid model, spontaneously broken supersymmetric (SB SUSY) theories \citep{hybrid_inflation1994_s4}, as shown in green in Figure~\ref{fig:ns_r}, is compatible with $\Lambda\mathrm{CDM}_\mathtt{TTBW}$, $\Lambda\mathrm{CDM}\xi_\mathtt{TTBW}$, $\Lambda$CDM$\xi_{\mathtt{TTBW}+\mathtt{SN}}$, and $\Lambda$CDM$\xi_{\mathtt{TTBW}+\tau_2}$. For all the above models, our results suggest a very small value of $r$ ($r\approx0$). We also show monomial potential models $V(\phi)\sim\phi^p$, with $p=$ 2, 1, and 2/3 as red lines ($50<N<60$) in Figure~\ref{fig:ns_r} for comparison \citep{monomial_inflation1983_s1, monomial_inflation2008_s2, monimial_inflation2010_s3, monimial_inflation2014_s4}.
\subsection{DESI BAO results}

\begin{figure*}
    \centering
    \includegraphics[width=0.55\linewidth]{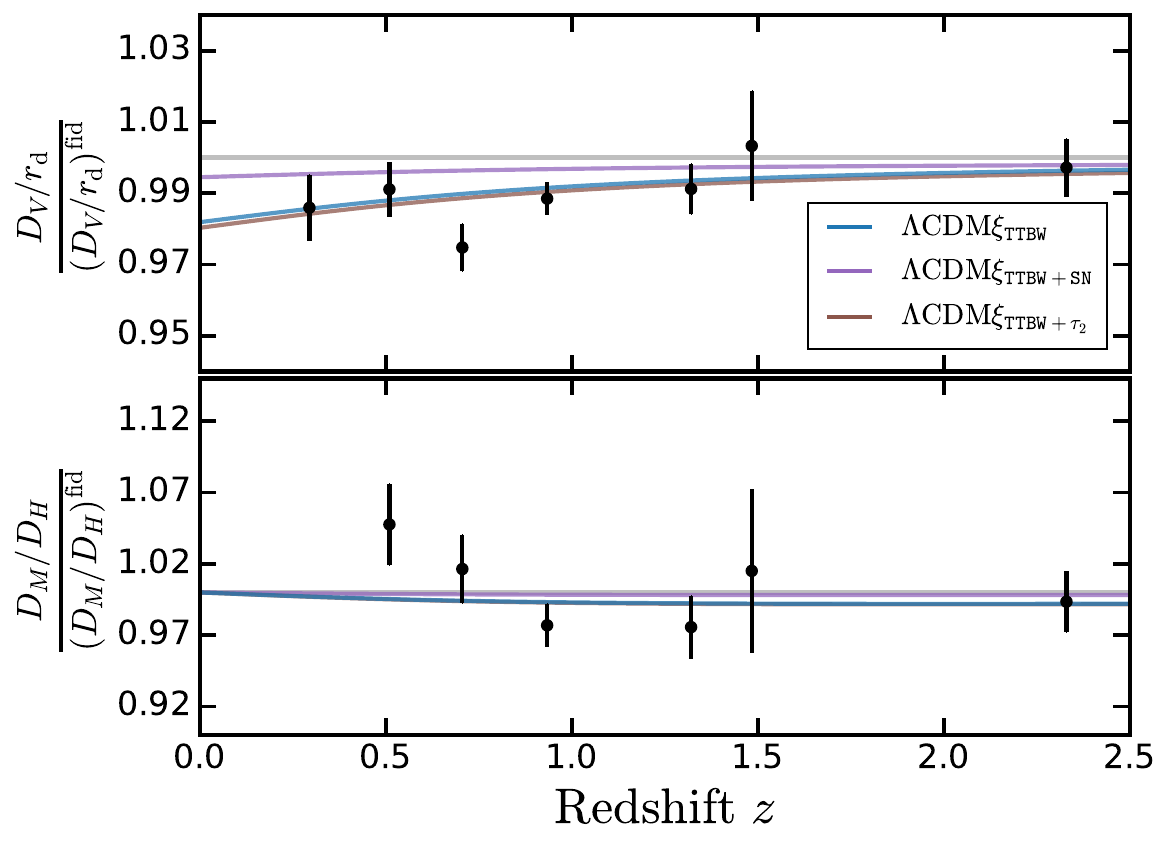}
    \caption{Ratios of $D_V/r_\mathrm{d}$ (upper panel) and $D_M/D_H$ (lower panel) to the corresponding predictions from $\Lambda$CDM$_\mathtt{P18}$, denoted as ``fid.'' Data with error bars are from the DESI BAO results \citep{desi2025arXiv250314745D,desi2025arXiv250314739D}. The blue, purple, and brown lines are calculated by the cosmological parameters from $\Lambda$CDM$\xi_\mathtt{TTBW}$, $\Lambda$CDM$\xi_{\mathtt{TTBW}+\mathtt{SN}}$, and $\Lambda$CDM$\xi_{\mathtt{TTBW}+\mathtt{\tau_2}}$, respectively.}
    \label{fig:desi_discrepancy}
\end{figure*}

As shown in Figure~\ref{fig:desi_discrepancy}, the recent DESI BAO measurements of $D_V/r_\mathrm{d}$ and $D_M/D_H$ deviate from the predictions of the $\Lambda$CDM model with Planck 2018 cosmological parameters ($\Lambda$CDM$_\mathtt{P18}$) \citep{desi2025arXiv250314745D, desi2025arXiv250314739D, desi2025arXiv250314738D}. Here, $D_M(z)$ ($D_H(z)$) is the transverse (line-of-sight) comoving distance, $D_V\equiv(zD_M^2D_H)^{1/3}$, and $r_\mathrm{d}$ is the sound horizon at the baryon drag epoch. 

As shown by colored lines in Figure~\ref{fig:desi_discrepancy}, the $\Lambda$CDM$\xi_\mathtt{TTBW}$ (as well as $\Lambda$CDM$\xi_{\mathtt{TTBW}+\mathtt{SN}}$ and $\Lambda$CDM$\xi_{\mathtt{TTBW}+\mathtt{\tau_2}}$) predictions agree better with the DESI BAO measurements than those from $\Lambda$CDM$_\mathtt{P18}$. This can be shown quantitatively by calculating
\begin{equation}
    \chi_\mathrm{DESI}^2=\sum_i\left(\frac{\alpha_i-\alpha^\mathrm{predict}_i}{\Delta\alpha_i}\right)^2,
\end{equation}
where $\alpha_i$ and $\Delta\alpha_i$ are $D_V/r_\mathrm{d}$ or $D_M/D_H$ and corresponding errors \citep{desi2025arXiv250314745D, desi2025arXiv250314739D, desi2025arXiv250314738D}, respectively, shown as black data points in Figure~\ref{fig:desi_discrepancy}, while the predictions on $\alpha_i$ based on $\Lambda$CDM$_\mathtt{P18}$, $\Lambda$CDM$\xi_\mathtt{TTBW}$, $\Lambda$CDM$\xi_{\mathtt{TTBW}+\mathtt{SN}}$, and $\Lambda$CDM$\xi_{\mathtt{TTBW}+\mathtt{\tau_2}}$ are denoted as $\alpha_i^\mathrm{predict}$. The summation is taken over all 13 data points. 

The resulting $\chi_\mathrm{DESI}^2$ values are 32.9, 12.2, 22.6, and 11.4 for $\Lambda$CDM$_\mathtt{P18}$, $\Lambda$CDM$\xi_\mathtt{TTBW}$, $\Lambda$CDM$\xi_{\mathtt{TTBW}+\mathtt{SN}}$, and $\Lambda$CDM$\xi_{\mathtt{TTBW}+\mathtt{\tau_2}}$, respectively. Using Akaike Information Criterion \citep[AIC;][]{AIC2007MNRAS, AICcavanaugh2019}, we obtain $\Delta \mathrm{AIC}$ = 16.7, 6.3, and 17.5 for $\Lambda$CDM$\xi_\mathtt{TTBW}$, $\Lambda$CDM$\xi_{\mathtt{TTBW}+\mathtt{SN}}$ and $\Lambda$CDM$\xi_{\mathtt{TTBW}+\mathtt{\tau_2}}$, relative to the $\Lambda$CDM$_\mathtt{P18}$. These values indicate a positive preference for $\Lambda$CDM$\xi_\mathtt{TTBW}$, $\Lambda$CDM$\xi_{\mathtt{TTBW}+\mathtt{SN}}$, and $\Lambda$CDM$\xi_{\mathtt{TTBW}+\mathtt{\tau_2}}$, based on the DESI data.

\section{Conclusion}\label{sc:conclusion}

In this Letter, we refit various cosmological data sets by the $\Lambda$CDM$\xi$ model, which introduces two additional neutrino parameters: the total mass $M_\nu$ and the degeneracy parameter $\xi_3$. We find that the $H_0$ and $S_8$ tensions are significantly alleviated only when the CMB polarization data are excluded ($\mathtt{TTBW}$), implying that these tensions exist between the CMB temperature and polarization data, both being global measurements.

Our main results can be summarized as follows:

\hspace*{\fill}

(i) For $\Lambda$CDM$\xi_\mathtt{TTBW}$, we find $\sim3\sigma$ evidence for both neutrino mass $M_\nu=0.57^{+0.17}_{-0.13}\,\mathrm{eV}$ and degeneracy parameter $\xi_3=1.13^{+0.41}_{-0.19}$ (68\% CL). Moreover, this $\mathcal{O}(1)$ neutrino degeneracy parameter is also consistent with the BBN data.

With a tighter prior on the optical depth $\tau$ ($\mathtt{TTBW}+\tau_2$), the constraints become $M_\nu=0.167^{+0.083}_{-0.097}\,\mathrm{eV}$ and $\xi_3=0.40^{+0.12}_{-0.40}$, with significance reduced to 1.8$\sigma$ and 0.5$\sigma$, respectively. This confirms that constraints on neutrino properties are sensitive to the adopted $\tau$ prior.

(ii) In the $\Lambda$CDM$\xi_\mathtt{TTBW}$ case, the scalar spectral index becomes consistent with, or slightly greater than, unity ($n_s = 1.019^{+0.022}_{-0.020}$). This opens the possibility for hybrid inflation models—including those predicting $n_s \gtrsim 1$—to remain viable. In particular, scenarios involving two scalar fields or SB SUSY inflation are compatible with our inferred $n_s$. By contrast, in the $\Lambda$CDM$\xi_{\mathtt{TTBW}+\tau_2}$ case, $n_s$ remains close to the standard Planck result ($n_s = 0.9777^{+0.0048}_{-0.0073}$).

(iii) The recent DESI BAO measurements of $D_V/r_\mathrm{d}$ and $D_M/D_H$ agree much better with the $\Lambda$CDM$\xi$ predictions than those of Planck $\Lambda$CDM. All three cases—$\Lambda$CDM$\xi_\mathtt{TTBW}$, $\Lambda$CDM$\xi_{\mathtt{TTBW}+\mathtt{SN}}$, and $\Lambda$CDM$\xi_{\mathtt{TTBW}+\tau_2}$—yield improved $\chi^2$ values and are preferred by AIC analysis.

\hspace*{\fill}

Upcoming CMB observations, such as those from the South Pole Telescope \citep{SPT2014}, ACT \citep{ACT2016}, Simons Observatory \citep{CMB_SO2019}, CMB-S4 \citep[Stage-4;][]{CMB_S42016_s1, CMB_S42019_s2}, the proposed CMB-HD \citep[Ultra-Deep, High-Resolution;][]{CMB_HD2019_s1, CMB_HD2019_s2, CMB_HD2020_s3}, and other future missions will provide further tests of this $\Lambda$CDM$\xi$ model and more precise measurements of the neutrino parameters.

\hspace*{\fill}

We thank HuanYuan Shan and Ji Yao for their assistance with the weak lensing data analysis as well as the anonymous referee for helpful comments. The computational resources used for the MCMCs in this work were kindly provided by the Chinese University of Hong Kong Central Research Computing Cluster. Furthermore, this research is supported by grants from the Research Grants Council of the Hong Kong Special Administrative Region, China, under Project No. AoE/P-404/18 and 14300223. All plots in this Letter are generated by \texttt{Getdist} \citep{getdist2019} and \texttt{Matplotlib} \citep{matlibplot2007}, and we also use \texttt{Scipy} \citep{scipy2020}, and \texttt{Numpy} \citep{numpy2020}.

\appendix
\section{impact of cosmological models on \texorpdfstring{$S_8$}{S8}}\label{sc:S8_MCMC}
Since the value of $S_8$ is model dependent \citep[e.g., Figure~13 in][]{DES_model2023prd}, we reanalyze the 2PCFs from the KiDS-1000 data \citep{asgari2021aa} using both the $\Lambda$CDM and $\Lambda$CDM$\xi$ models, implemented via the \texttt{CCL}\footnote{\hyperlink{https://github.com/LSSTDESC/CCL}{https://github.com/LSSTDESC/CCL}} \citep{ccl2019ApJS} with the Takahashi halofit model \citep{Takahashi2012ApJ}, following the approach of \cite{yao2023A&A}. We adopted the same priors for the cosmological and nuisance parameters as detailed in Table~2 of \cite{asgari2021aa}, with uniform priors of $[0,3]$ eV for the neutrino mass $M_\nu$ and $[0,2]$ for the degeneracy parameter $\xi_3$.

Our analysis yields $S_8 = 0.759^{+0.021}_{-0.025}$ for the $\Lambda$CDM model and $S_8 = 0.748^{+0.019}_{-0.022}$ for the $\Lambda$CDM$\xi$ model. The $S_8$ value from the $\Lambda$CDM analysis is consistent with the KiDS-1000 results \citep{asgari2021aa}, whereas the result from the $\Lambda$CDM$\xi$ model is slightly lower.

One caveat is that the Takahashi halofit model does not have an explicit neutrino dependence. We assume that the nonlinearity remains valid for neutrinos with finite masses ($M_\nu$) and degeneracy parameters ($\xi_i$). A fully consistent investigation of this assumption is left for future work.

\section{Full parameter table and contours}\label{sc:cosmo_tables}

\begin{figure*}
    \centering
    \includegraphics[width=\linewidth]{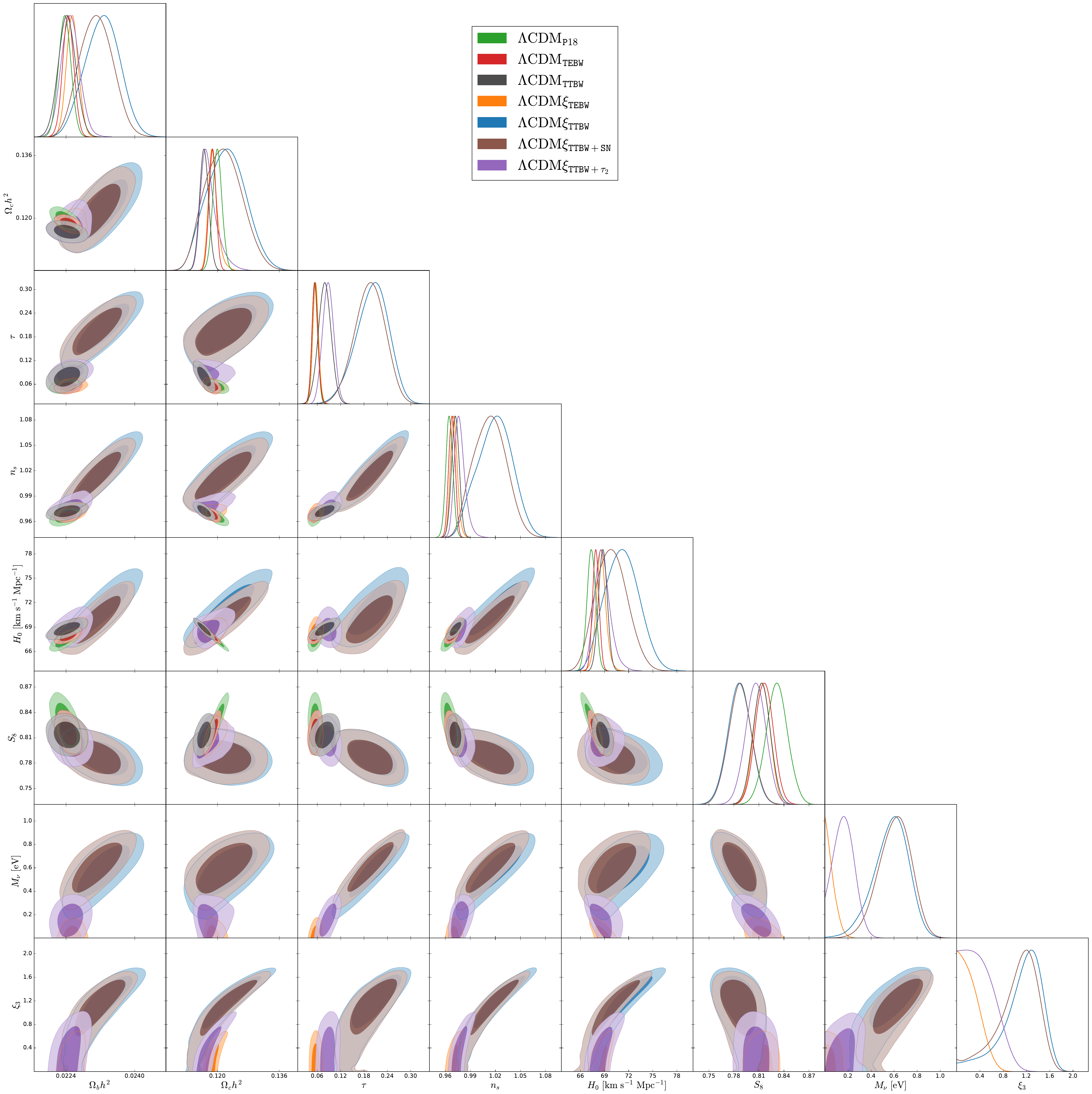}
    \caption{1D marginalized posterior PDFs and 2D contours (68\% and 95\% CL) of cosmological parameters for different models from different data sets as shown in the legend.}
    \label{fig:full_CMB_contours}
\end{figure*}

Table~\ref{tab:full_cosmo} shows the mean values of the cosmological parameters with 68\% CL from $\Lambda\mathrm{CDM}_\mathtt{P18}$, $\Lambda\mathrm{CDM}_\mathtt{TEBW}$, $\Lambda\mathrm{CDM}_\mathtt{TTBW}$, $\Lambda\mathrm{CDM}\xi_\mathtt{TEBW}$, $\Lambda\mathrm{CDM}\xi_\mathtt{TTBW}$, $\Lambda\mathrm{CDM}\xi_{\mathtt{TTBW}+\mathtt{SN}}$, and $\Lambda\mathrm{CDM}\xi_{\mathtt{TTBW}+\tau_2}$, and we follow the definitions of the cosmological parameters as shown in Table~1 of \cite{planck2014aa}. The first group shows the free parameters used in each case, while the second group denotes the derived parameters based on the first group. Figure~\ref{fig:full_CMB_contours} shows the 2D contours for the combination of models and data sets listed in Table~\ref{tab:full_cosmo}.

\begin{table*}[htbp]
\centering
\rotatebox{90}{
  \begin{minipage}{\textheight} 
    \centering
    \begin{tabular}{ccclcccl}
    \hline
    \hline
    Parameter & $\Lambda\mathrm{CDM}_\mathtt{P18}$ & $\Lambda\mathrm{CDM}_\mathtt{TEBW}$ &$\Lambda\mathrm{CDM}_\mathtt{TTBW}$  & $\Lambda\mathrm{CDM}\xi_\mathtt{TEBW}$  & $\Lambda\mathrm{CDM}\xi_\mathtt{TTBW}$ & $\Lambda\mathrm{CDM}\xi_\mathtt{TTBW+SN}$ & $\Lambda\mathrm{CDM}\xi_\mathtt{TTBW}+\tau_2$\\
    \hline     
    $\Omega_bh^2$& $0.02237\pm0.00014$&  $0.02251\pm0.00013$ &$0.02242\pm0.00020$ & $0.02255^{+0.00013}_{-0.00015}$ & $0.02325^{+0.00042}_{-0.00039}$& $0.02311^{+0.00038}_{-0.00039}$& $0.02250^{+0.00020}_{-0.00022}$\\
    $\Omega_ch^2$& $0.1200\pm0.0012$&  $0.11818^{+0.00083}_{-0.00084}$ &$0.1166\pm0.0012$ & $0.1190^{+0.00091}_{-0.0014}$ & $0.1223^{+0.0050}_{-0.0051}$& $0.1216^{+0.0044}_{-0.0052}$& $0.1180^{+0.0012}_{-0.0026}$\\
    $100\theta_\mathrm{MC}$&  $1.04091^{+0.00031}_{-0.00030}$&  $1.04109^{+0.00028}_{-0.00029}$ &$1.04119^{+0.00042}_{-0.00041}$ & $1.04101\pm0.00031$ & $1.04056\pm0.00061$& $1.04060^{+0.00061}_{-0.00060}$& $1.04103\pm0.00046$\\
    $\tau$&  $0.0542^{+0.0073}_{-0.0072}$&  $0.0568^{+0.0070}_{-0.0075}$ &$0.079\pm0.016$ & $0.0562^{+0.0070}_{-0.0077}$ & $0.201^{+0.045}_{-0.037}$& $0.195^{+0.041}_{-0.037}$& $0.088\pm0.014$\\
    $\ln(10^{10}A_s)$&  $3.044\pm0.014$&  $3.045\pm0.014$ &$3.085\pm0.030$ & $3.046^{+0.014}_{-0.015}$  & $3.337^{+0.096}_{-0.077}$& $3.324^{+0.086}_{-0.078}$& $3.107^{+0.027}_{-0.030}$\\
    $n_s$ & $0.9649^{+0.0042}_{-0.0041}$& $0.9688\pm0.0037$ &$0.9721^{+0.0046}_{-0.0045}$ & $0.9704^{+0.0038}_{-0.0044}$ & $1.019^{+0.022}_{-0.020}$& $1.013\pm0.020$& $0.9777^{+0.0048}_{-0.0073}$\\
    $M_\nu$ [eV]&  --- &  --- &---  & $0.055^{+0.013}_{-0.055}$ & $0.57^{+0.17}_{-0.13}$& $0.61^{+0.15}_{-0.13}$& $0.167^{+0.083}_{-0.097}$\\
    $\xi_3$ &  --- &  --- &---  & $0.246^{+0.071}_{-0.25}$& $1.13^{+0.41}_{-0.19}$& $1.05^{+0.42}_{-0.21}$& $0.40^{+0.12}_{-0.40}$\\
    \hline
    $H_0\,[\mathrm{km}\,\mathrm{s}^{-1}\,\mathrm{Mpc}^{-1}]$&  $67.35^{+0.54}_{-0.53}$&  $68.17\pm0.38$ &$68.79\pm0.55$ & $68.57^{+0.56}_{-0.62}$ & $71.2\pm2.1$& $69.9^{+1.8}_{-2.1}$& $68.75^{+0.82}_{-1.1}$\\
    $\Omega_\Lambda$&  $0.6847^{+0.0076}_{-0.0070}$&  $0.6958^{+0.0050}_{-0.0049}$ &$0.7048\pm0.0068$ & $0.6977^{+0.0068}_{-0.0057}$ & $0.6974\pm0.0085$& $0.6872^{+0.0079}_{-0.0077}$& $0.6987^{+0.0090}_{-0.0082}$\\
    $\Omega_m$&  $0.3153^{+0.0070}_{-0.0076}$&  $0.3042^{+0.0049}_{-0.0050}$ &$0.2952\pm0.0068$ & $0.3023^{+0.0057}_{-0.0068}$ & $0.3026\pm0.0085$& $0.3128^{+0.0077}_{-0.0079}$& $0.3013^{+0.0082}_{-0.0090}$\\
    $\Omega_mh^2$&  $0.1430\pm0.0011$&  $0.14134\pm0.00080$ &$0.1396\pm0.0011$ & $0.14210^{+0.00093}_{-0.0015}$& $0.1534^{+0.0075}_{-0.0079}$& $0.1528^{+0.0063}_{-0.0081}$& $0.1423^{+0.0012}_{-0.0030}$\\
    $\Omega_mh^3$&  $0.09632^{+0.00031}_{-0.00028}$&  $0.09634^{+0.00030}_{-0.00029}$ &$0.09605\pm0.00045$ & $0.09744^{+0.00039}_{-0.0013}$& $0.1093^{+0.0083}_{-0.0087}$& $0.1069^{+0.0066}_{-0.0091}$& $0.09787^{+0.00088}_{-0.0030}$\\
    $\sigma_8$&  $0.8111^{+0.0061}_{-0.0059}$&  $0.8062\pm0.0055$ &$0.820\pm0.011$ & $0.8104^{+0.012}_{-0.0077}$& $0.783\pm0.017$& $0.771\pm0.016$& $0.805^{+0.018}_{-0.016}$\\
    $S_8\equiv\sigma_8(\Omega_m/0.3)^{0.5}$&  $0.832\pm0.013$&  $0.8118^{+0.0090}_{-0.0092}$ &$0.813\pm0.010$ & $0.8134^{+0.0098}_{-0.0097}$ & $0.787\pm0.014$& $0.787\pm0.014$& $0.806\pm0.012$\\
    $\sigma_8\Omega_m^{0.25}$&  $0.6078^{+0.0065}_{-0.0064}$&  $0.5988\pm0.0050$ &$0.6044\pm0.0069$ & $0.6009^{+0.0073}_{-0.0060}$ & $0.581\pm0.011$& $0.577\pm0.010$& $0.5961^{+0.011}_{-0.0095}$\\
    $z_{re}$&  $7.66^{+0.73}_{-0.72}$&  $7.86^{+0.72}_{-0.70}$ &$9.9^{+1.5}_{-1.3}$ & $7.81^{+0.73}_{-0.72}$ & $19.3^{+3.3}_{-2.5}$& $19.0^{+3.0}_{-2.5}$& $10.8\pm1.2$\\
    $10^9A_s$&  $2.100\pm0.030$&  $2.102^{+0.028}_{-0.031}$ &$2.187\pm0.066$ & $2.104^{+0.029}_{-0.032}$ & $2.82^{+0.24}_{-0.25}$& $2.78\pm0.23$& $2.236^{+0.059}_{-0.068}$\\
    $10^9A_se^{-2\tau}$&  $1.884\pm0.011$&  $1.876\pm0.010$ &$1.866^{+0.012}_{-0.011}$ & $1.880^{+0.011}_{-0.012}$ & $1.882^{+0.021}_{-0.018}$& $1.880\pm0.020$& $1.874^{+0.012}_{-0.016}$\\
    Age [Gyr]&  $13.798^{+0.023}_{-0.024}$&  $13.771\pm0.019$ &$13.764\pm0.029$ & $13.715^{+0.069}_{-0.030}$ & $13.22^{+0.32}_{-0.37}$& $13.34^{+0.35}_{-0.33}$& $13.695^{+0.15}_{-0.058}$\\
    $z_*$&  $1089.92^{+0.26}_{-0.25}$&  $1089.59\pm0.20$ &$1089.55\pm0.30$ & $1089.65\pm0.22$ & $1089.99^{+0.56}_{-0.57}$& $1090.01^{+0.55}_{-0.56}$& $1089.71^{+0.30}_{-0.37}$\\
    $r_*$ [Mpc]&  $144.43^{+0.26}_{-0.27}$&  $144.80^{+0.20}_{-0.21}$ &$145.29^{+0.29}_{-0.30}$ & $144.33^{+0.67}_{-0.21}$ & $139.1^{+3.3}_{-3.9}$& $139.8^{+3.6}_{-3.4}$& $144.22^{+1.4}_{-0.42}$\\
    $100\theta_*$&  $1.04109^{+0.00030}_{-0.00029}$&  $1.04126^{+0.00028}_{-0.00029}$ &$1.04140^{+0.00041}_{-0.00040}$ & $1.04117^{+0.00034}_{-0.00030}$ & $1.04044\pm0.00080$& $1.04056^{+0.00080}_{-0.00078}$& $1.04120^{+0.00054}_{-0.00046}$\\
    $z_\mathrm{drag}$&  $1059.93^{+0.30}_{-0.31}$&  $1060.12\pm0.29$ &$1059.80^{+0.44}_{-0.45}$ & $1060.31^{+0.31}_{-0.39}$ & $1062.9^{+1.6}_{-1.5}$& $1062.4^{+1.5}_{-1.6}$& $1060.21^{+0.48}_{-0.70}$\\
    $r_\mathrm{drag}$ [Mpc]&  $147.09\pm0.26$&  $147.42^{+0.21}_{-0.22}$ &$147.96\pm0.32$ & $146.93^{+0.70}_{-0.22}$ & $141.4^{+3.4}_{-4.0}$& $142.2^{+3.8}_{-3.5}$& $146.84^{+1.5}_{-0.45}$\\
    $k_\mathrm{D}\,\mathrm{[Mpc^{-1}]}$&  $0.14086\pm0.00030$&  $0.14062^{+0.00028}_{-0.00027}$ &$0.13999\pm0.00042$ & $0.14099^{+0.00029}_{-0.00058}$ & $0.1448^{+0.0029}_{-0.0028}$& $0.1442^{+0.0025}_{-0.0030}$& $0.14077^{+0.00048}_{-0.0012}$\\
    $z_\mathrm{eq}$&  $3402^{+27}_{-26}$&  $3362\pm19$ &$3321^{+27}_{-26}$ & $3382^{+20}_{-34}$ & $3479^{+126}_{-131}$& $3459^{+111}_{-132}$& $3357^{+29}_{-64}$\\
    $k_\mathrm{eq}\,\mathrm{[Mpc^{-1}]}$&  $0.010384^{+0.000081}_{-0.000079}$&  $0.010262\pm0.000058$ &$0.010137\pm0.000081$ & $0.010338^{+0.000061}_{-0.00012}$& $0.01093^{+0.00055}_{-0.00058}$& $0.01083^{+0.00047}_{-0.00059}$& $0.010290^{+0.000091}_{-0.00025}$\\
    $100\theta_{\mathrm{s,eq}}$&  $0.4494^{+0.0025}_{-0.0026}$&  $0.4533^{+0.0019}_{-0.0018}$ &$0.4572\pm0.0026$ & $0.4512^{+0.0035}_{-0.0020}$ & $0.439^{+0.013}_{-0.014}$& $0.441^{+0.014}_{-0.013}$& $0.4532^{+0.0070}_{-0.0030}$\\
    \hline\hline
    \end{tabular}
    \caption{Mean values of cosmological parameters (68\% CL) for $\Lambda\mathrm{CDM}_\mathtt{P18}$, $\Lambda\mathrm{CDM}_\mathtt{TEBW}$ $\Lambda\mathrm{CDM}_\mathtt{TTBW}$, $\Lambda\mathrm{CDM}\xi_\mathtt{TEBW}$, $\Lambda\mathrm{CDM}\xi_\mathtt{TTBW}$, $\Lambda\mathrm{CDM}\xi_{\mathtt{TTBW}+\mathtt{SN}}$, and $\Lambda\mathrm{CDM}\xi_{\mathtt{TTBW}+\tau_2}$. The first group lists free parameters used in each case, while the second group shows the derived parameters based on the first group.}
    \label{tab:full_cosmo}
  \end{minipage}%
}
\end{table*}

\section{Robustness of fitting results}\label{sc:robustness}

In this section, we investigate the robustness of our main conclusion by separately updating the Planck, BAO, and DES data sets. In each case, we find that the $\Lambda$CDM$\xi$ model significantly alleviates the $H_0$ and $S_8$ tensions, as shown in Table~\ref{tab:robustness}.\footnote{Except for $S_8$ in the cases of $\mathtt{TTBW}_\mathrm{DESY3}$ and $\mathtt{TTBW}_\mathrm{PR4}$, where only a slight alleviation is observed.}

(i) Planck 2020 data: We update the Planck 2018 high-multipole temperature TT and CMB lensing data with Planck 2020 PR4 \citep{PR42024}, denoted as $\mathtt{TTBW}_\mathrm{PR4}$, which combines Planck 2020 PR4 \texttt{HiLLiPoP} TT and CMB lensing data with Planck 2018 lowT data, along with BAO and DES measurements. As shown in Table~\ref{tab:robustness}, the $\Lambda$CDM$\xi$ model does not suffer from the $H_0$ tension, and the $S_8$ tension is slightly alleviated when fitting the $\mathtt{TTBW}_\mathrm{PR4}$ data.

(ii) DESI BAO data: We substitute the BAO data in the $\mathtt{TTBW}$ data set with the latest DESI BAO measurements \citep{desi2025arXiv250314745D, desi2025arXiv250314739D, desi2025arXiv250314738D}, denoted as $\mathtt{TTBW}_\mathrm{DESI}$. As shown in Table~\ref{tab:robustness}, the $H_0$ and $S_8$ tensions are significantly alleviated.

(iii) DES Y3 data: Finally, we replace the DES Y1 data in the $\mathtt{TTBW}$ data set with the DES Y3 data \citep{DES2022}, denoted as $\mathtt{TTBW}_\mathrm{DESY3}$. As shown in Table~\ref{tab:robustness}, $\Lambda$CDM$\xi_{\mathtt{TTBW}_\mathrm{DESY3}}$ again alleviates the $H_0$ and $S_8$ tensions.

In all three cases, the $\Lambda$CDM$\xi$ model remains consistent with alleviating the tensions, demonstrating the robustness of our findings across different updated data.

\begin{table*}
    \centering
    \renewcommand{\arraystretch}{1.3}
    \setlength{\tabcolsep}{5pt}      
    \begin{tabular}{lcccc}
    \hline\hline
         & \multicolumn{2}{c}{$\Lambda\mathrm{CDM}$} & \multicolumn{2}{c}{$\Lambda\mathrm{CDM}\xi$} \\
         & $H_0\,[\mathrm{km}\,\mathrm{s}^{-1}\mathrm{Mpc}^{-1}]$ & $S_8$ 
         & $H_0\,[\mathrm{km}\,\mathrm{s}^{-1}\mathrm{Mpc}^{-1}]$ & $S_8$ \\
    \hline
    \multirow{2}{*}{$\mathtt{TTBW}_\mathrm{PR4}$} 
         & $68.63 \pm 0.53$   & $0.812 \pm 0.010$  
         & $70.9^{+1.7}_{-2.3}$ & $0.792 \pm 0.013$ \\
         & 3.8               & 2.1 
         & 1.0               & 1.8\\
    \hline
    \multirow{2}{*}{$\mathtt{TTBW}_\mathrm{DESI}$}
         & $68.86^{+0.34}_{-0.33}$& $0.8129\pm0.0096$& $72.0^{+2.0}_{-1.9}$& $0.787 \pm 0.014$\\
         & 3.8& 2.2& 0.4& 1.6\\
    \hline
    \multirow{2}{*}{$\mathtt{TTBW}_\mathrm{DESY3}$}& $67.87^{+0.42}_{-0.41}$ & $0.8120^{+0.0097}_{-0.0098}$  
         & $69.8^{+1.4}_{-2.1}$& $0.798^{+0.016}_{-0.014}$\\
         & 4.6               & 2.1& 1.6& 1.9\\
    \hline\hline
    \end{tabular}
    \caption{Mean values (68\% CL) of $H_0$ (in $\mathrm{km}\,\mathrm{s}^{-1}\mathrm{Mpc}^{-1}$) and $S_8$ from different data sets ($\mathtt{TTBW}_\mathrm{PR4}$, $\mathtt{TTBW}_\mathrm{DESI}$, and $\mathtt{TTBW}_\mathrm{DESY3}$) and models ($\Lambda\mathrm{CDM}$ and $\Lambda\mathrm{CDM}\xi$), followed by their respective deviations, $n_\sigma$ with respect to the local measurements.}
    \label{tab:robustness}
\end{table*}

\section{Consistency with BBN}\label{sc:BBN_He3}

The BBN observables, $Y_\mathrm{P}$ (the helium-4 mass fraction) and $\mathrm{D}/\mathrm{H}$ (the deuterium-to-hydrogen ratio) can remain unchanged despite the larger $\Omega_bh^2$ from $\Lambda$CDM$\xi_\mathtt{TTBW}$ results ($\Delta\Omega_bh^2=0.001$) if $\Delta N_\nu$ and $\xi_{\nu_e}$ are also increased ($\Delta N_\nu=0.71$, $\xi_{\nu_e}=0.04$). Here, we check whether the helium-3-to-hydrogen ratio, $\prescript{3}{}{\mathrm{He}}/\mathrm{H}$, is affected by such $\Delta N_\nu$ and $\xi_{\nu_e}$, using the following fitting formula \citep{pitrou2018bbn},
\begin{equation}\label{eq:BBN_estimator_He3}
    \frac{ \Delta \prescript{3}{}{ \mathrm{He} }/\mathrm{H} }
         { \overline{\prescript{3}{}{\mathrm{He}}/\mathrm{H}} } 
         = 
          -0.57\frac{\Delta \Omega_b h^2}{\overline{\Omega_b h^2}} 
          + 0.14\frac{\Delta N_\nu}{3.0} 
          - 0.18\xi_{\nu_e}.
\end{equation}
We obtain $
            \frac{ \Delta \prescript{3}{}{ \mathrm{He} }/\mathrm{H} }
                 { \overline{\prescript{3}{}{\mathrm{He}}/\mathrm{H}} } = 0.00032$, 
which is consistent with 0 within the uncertainty. Consequently, the value of $\xi_3\sim\mathcal{O}(1)$ is consistent with major BBN observations.

\section{
Semianalytic model on 
\texorpdfstring{$H_0$}{H0}
}\label{sc:semi_analysis}
\subsection{
Perturbation analysis of 
\texorpdfstring{$\theta_*$}{thetastar}
}\label{sc:semi_analysis_theta}

In this section, following \cite{Yeung_2021}, we use a semianalytic approach to understand the neutrino effects on $H_0$. We start from the sound horizon angular size ($\theta_*$) at the epoch of photon decoupling $z_*$, 
\begin{equation}
    \theta_* = \frac{r_s}{D_M},
\end{equation}
where $r_s$ and $D_M$ are the comoving sound horizon and distance to the last scattering, respectively:
\begin{equation}
    \begin{aligned}
        r_s(z_*) &= \int_{z_*}^\infty \frac{dz}{\sqrt{3(1+R)}H(z)}, \\
        D_M(z_*) &= \int_0^{z_*} \frac{dz}{H(z)},
    \end{aligned}
\end{equation}
where $R\equiv 3\rho_b/4\rho_\gamma$ is the ratio between baryon and photon energy densities. 

For convenience, we take $c=\hbar=k_B=1$ and define the following kernels:
\begin{equation}
    \begin{aligned}
        G(x,\xi_i) &\equiv \frac{1}{e^{x-\xi_i}+1} + \frac{1}{e^{x+\xi_i}+1}, \\
        F(x,\xi_i) &\equiv \frac{1}{e^{x-\xi_i}+1} - \frac{1}{e^{x+\xi_i}+1}, \\
        K_1(x,y)   &\equiv 2x\sqrt{x^2+y^2} + \frac{x^3}{\sqrt{x^2+y^2}}, \\
        K_2(x,y)   &\equiv 2x\sqrt{x^2+y^2} + \frac{5x^2}{\sqrt{x^2+y^2}} 
                  - \frac{x^4}{(x^2+y^2)^{3/2}}, 
    \end{aligned}
\end{equation}
and functionals
\begin{equation}
    \begin{aligned}
        \mathcal{I}_r[f(z)] &\equiv \int_{z_*}^\infty \frac{f(z)}{[E^\mathrm{ref}(z)]^3}\frac{dz}{\sqrt{3(1+R)}}, \\
        \mathcal{I}_D[f(z)] &\equiv \int_0^{z_*} \frac{f(z)}{[E^\mathrm{ref}(z)]^3}dz,
    \end{aligned}
\end{equation}
where $E^\mathrm{ref}(z)\equiv H^\mathrm{ref}(z)/H_0^\mathrm{ref}$.

As $\theta_*$ is well constrained by CMB data, we expand $r_s$ and $D_M$ around reference values of $M_\nu$ and $H_0$, denoted as $M_\nu^\mathrm{ref}$ and $H_0^\mathrm{ref}$, to the lowest order by fixing $\theta_*$, $\xi_i$, and other cosmological parameters ($\Omega_b h^2, \Omega_c h^2, \Omega_\gamma h^2$). We obtain
\begin{equation}
    C_H \frac{H_0 - H_0^\mathrm{ref}}{H_0^\mathrm{ref}} = C_m \frac{M_\nu - M_\nu^\mathrm{ref}}{M_\nu^\mathrm{ref}},
\end{equation}
where 
\begin{equation}\label{eq:taylor_CH_Cm}
\resizebox{\columnwidth}{!} 
{$
    \begin{aligned}
              C_H &= \mathcal{I}_r(1) - \theta_*\mathcal{I}_D(1), \\
              C_m &= \theta_* 
                     \mathcal{I}_D[C_{\rho m}(z)- (1+z)^3C_{\rho m}(0)] \\
                  &- \mathcal{I}_r[C_{\rho m}(z) - (1+z)^3C_{\rho m}(0)], \\
    C_{\rho m}(z) &= \frac{T_{\nu0}^4(1+z)^2}{4\uppi^2\rho_{\mathrm{crit},0}^\mathrm{ref}}
                     \int_0^\infty dx \sum_i G(x,\xi_i)
                     \frac{x^2\tilde m^2_i}{\sqrt{x^2+(\tilde m_i/(1+z))^2}},
    \end{aligned}    
$}
\end{equation}
and $\tilde m_i\equiv m_i/T_{\nu0}$. $T_{\nu0}$ and $\rho_{\mathrm{crit},0}$ are the relic neutrino temperature and critical energy density of the Universe today, respectively.

Similarly, we expand around $\xi_i^\mathrm{ref}$ ($i=1,2,3$) and $H_0^\mathrm{ref}$ to investigate the correlation between $\xi_i$ and $H_0$, while for $\xi_i$, we keep up to the second-order term,
\begin{equation}
    C_H \frac{H_0 - H_0^\mathrm{ref}}{H_0^\mathrm{ref}} 
    = 
    \sum_i C_{\xi,i}^1 (\xi_i - \xi_i^\mathrm{ref})
    + \sum_i C_{\xi,i}^2 (\xi_i - \xi_i^\mathrm{ref})^2,
\end{equation}
where
\begin{equation}\label{eq:taylor_Cxi}
    \begin{aligned}
        C_{\xi,i}^1 &= \theta_* 
                       \mathcal{I}_D[C_{\rho\xi,i}^1(z) - (1+z)^3C_{\rho\xi,i}^1(0)] \\
                    &- \mathcal{I}_r[C_{\rho\xi,i}^1(z) - (1+z)^3C_{\rho\xi,i}^1(0)], \\
        C_{\xi,i}^2 &= \theta_* 
                       \mathcal{I}_D[C_{\rho\xi,i}^2(z) - (1+z)^3C_{\rho\xi,i}^2(0)] \\
                    &- \mathcal{I}_r[C_{\rho\xi,i}^2(z) - (1+z)^3C_{\rho\xi,i}^2(0)], \\
        C_{\rho\xi,i}^1(z) &= 
        \frac{T_{\nu0}^4(1+z)^4}{4\uppi^2\rho_{\mathrm{crit},0}^\mathrm{ref}}
        \int_0^\infty dx K_1(x,\tilde m_i/(1+z)) F(x,\xi_i) ,\\
        C_{\rho\xi,i}^2(z) &= 
        \frac{T_{\nu0}^4(1+z)^4}{8\uppi^2\rho_{\mathrm{crit},0}^\mathrm{ref}}
        \int_0^\infty dx K_2(x,\tilde m_i/(1+z)) G(x,\xi_i).
    \end{aligned}
\end{equation}
We can find from the above that, when $\xi_i$ approaches zero, $F(x,\xi_i)\rightarrow0$; then, $C_{\rho\xi,i}^1$ also approaches zero, causing the first-order coefficient ($C_{\xi,i}^1$) to become negligible and allowing the second-order perturbation term to dominate. Therefore, unlike $M_\nu$, it is necessary to include terms up to second order in $\xi_i$.

\begin{table}[t]
\centering
\begin{tabular}{lccc}
\hline\hline
 & $\xi_3^\mathrm{ref}=0.2$ & $\xi_3^\mathrm{ref}=1.1$ & $\xi_3^\mathrm{ref}=1.6$ \\
 & $M_\nu^\mathrm{ref}=0.33$ & $M_\nu^\mathrm{ref}=0.55$ & $M_\nu^\mathrm{ref}=0.77$ \\
 & $H_0^\mathrm{ref}=67.7$ & $H_0^\mathrm{ref}=70.6$ & $H_0^\mathrm{ref}=74.5$ \\
\hline
$C_m/C_H$ & -0.026 & -0.045 & -0.067 \\
$C_{\xi,1}^1/C_H$ & -0.006 & -0.036 & -0.051 \\
$C_{\xi,2}^1/C_H$ & 0.012 & 0.069 & 0.099 \\
$C_{\xi,3}^1/C_H$ & 0.029 & 0.163 & 0.236 \\
$C_{\xi,1}^2/C_H$ & 0.073 & 0.067 & 0.062 \\
$C_{\xi,2}^2/C_H$ & 0.074 & 0.071 & 0.070 \\
$C_{\xi,3}^2/C_H$ & 0.074 & 0.091 & 0.105 \\
\hline\hline
\end{tabular}
\caption{Selected expansion reference points and calculated coefficients of Eqs.~(\ref{eq:taylor_CH_Cm}) and (\ref{eq:taylor_Cxi}). Here, $M_\nu^\mathrm{ref}$ and $H_0^\mathrm{ref}$ are given in eV and $\mathrm{km}\,\mathrm{s}^{-1}\,\mathrm{Mpc}^{-1}$, respectively.}
\label{tab:taylor_calcoe}
\end{table}

We choose three $\xi_3^\mathrm{ref}$ points and obtain the corresponding $M_\nu^\mathrm{ref}$ and $H_0^\mathrm{ref}$ from the mean values in MCMC chains. The reference values for $\xi_3^\mathrm{ref}$, $M_\nu^\mathrm{ref}$, and $H_0^\mathrm{ref}$ are listed in Table~\ref{tab:taylor_calcoe}. Then, the corresponding coefficients ($C_m/C_H$, $C_{\xi,i}^1/C_H$, and $C_{\xi,i}^2/C_H$) can be calculated according to Eqs. (\ref{eq:taylor_CH_Cm}) and (\ref{eq:taylor_Cxi}), which are also summarized in Table~\ref{tab:taylor_calcoe}.

From the right panels of Figure~\ref{fig:taylor_results} in Section~\ref{sc:Mnu_xi3_H0}, we find that our perturbation analyses agree well with the MCMC results.

\subsection{BAO constraints}\label{sc:semi_analysis_BAO}

\begin{table*}
    \centering
    \renewcommand{\arraystretch}{1.3} 
    \setlength{\tabcolsep}{4pt} 
    \small 
    \begin{tabular}{lccccc}
    \hline
    \hline
         & \textbf{$z_1$} & \textbf{$z_2$} & \textbf{$z_3$} & \textbf{$z_4$} & \textbf{$z_5$} \\
    \hline
    Redshift \( z \) & 0.38 & 0.51 & 0.61 & 0.106 & 0.15 \\
    \hline
    \textbf{$r_{\mathrm{d,fid}}D_M(z)/r_\mathrm{drag}(z)$} [Mpc] & 1512.39 & 1975.22 & 2306.68 & --- & --- \\
    \textbf{$H(z)r_\mathrm{drag}(z)/r_{\mathrm{d,fid}}$} [\(\mathrm{km}\,\mathrm{s}^{-1}\,\mathrm{Mpc}^{-1}\)] & 81.2087 & 90.9029 & 98.9647 & --- & --- \\
    \hline
    \textbf{$r_\mathrm{drag}(z)/D_V(z)$} & --- & --- & --- & 0.336 $\pm$ 0.015 & --- \\
    \textbf{$D_V(z)/r_\mathrm{drag}(z)$} & --- & --- & --- & --- & 4.47 $\pm$ 0.17 \\
    \hline
    \hline
    \end{tabular}
    \caption{BAO measurements at various redshifts. The first group is from \cite{bao2017mnras} with fiducial \(r_{d,\mathrm{fid}} = 147.78\,\mathrm{Mpc}\), while the subsequent data sets are from \cite{bao2011mnras} and \cite{bao2015mnras}, respectively.}
    \label{tab:BAO_data}
\end{table*}

Based on the derived relation between ($M_\nu$,$H_0$) or ($\xi_3$,$H_0$), we can estimate the constraints from BAO data. BAO data used in Planck 2018 \citep{bao2011mnras,bao2015mnras,bao2017mnras} are summarized in Table~\ref{tab:BAO_data}, where $r_\mathrm{drag}$ is the comoving sound horizon at $z_\mathrm{drag}$,
\begin{equation}
    r_\mathrm{drag} = \int_{z_\mathrm{drag}}^\infty \frac{dz}{\sqrt{3(1+R)}H(z)},
\end{equation}
and the spherically averaged distance $D_V$ is
\begin{equation}
    D_V(z) \equiv \left[zD_M^2(z)D_H(z)\right]^{1/3},
\end{equation}
where $D_H\equiv c/H(z)$.

We define the quantities used in \cite{bao2017mnras} as $\boldsymbol{d}$,
\begin{equation}
    \begin{aligned}
        \boldsymbol{d} \equiv 
        &\left(r_{\mathrm{d,fid}}\frac{D_M(z_1)}{r_\mathrm{drag}}, 
        H(z_1)\frac{r_\mathrm{drag}}{r_{\mathrm{d,fid}}},\right.\\
        &\quad r_{\mathrm{d,fid}}\frac{D_M(z_2)}{r_\mathrm{drag}},
         H(z_2)\frac{r_\mathrm{drag}}{r_{\mathrm{d,fid}}},\\
        &\quad\left.r_{\mathrm{d,fid}}\frac{D_M(z_3)}{r_\mathrm{drag}}, 
         H(z_3)\frac{r_\mathrm{drag}}{r_{\mathrm{d,fid}}}
        \right).
    \end{aligned}
\end{equation}

Then we define $\chi^2$ as
\begin{equation}
    \begin{aligned}
        \chi^2
        &\equiv
        \sum_{ij}(\boldsymbol{d}_i-\boldsymbol{d}_i^\mathrm{obv})
                  C^{-1}_{ij}
                 (\boldsymbol{d}_j-\boldsymbol{d}_i^\mathrm{obv})^T \\
        &+ \left[\frac{r_\mathrm{drag}/D_V(z_4)-0.336}{0.015}\right]^2\\
        &+ \left[\frac{r_\mathrm{drag}/D_V(z_5)-4.47}{0.17}\right]^2,        
    \end{aligned}
\end{equation}
where $\boldsymbol{d}_i^\mathrm{obv}$ is the corresponding observation data summarized in Table~\ref{tab:BAO_data} and $C_{ij}$ is the covariance matrix between the BAO measurements \citep{bao2017mnras}:
\begin{equation}
\resizebox{0.88\columnwidth}{!} 
{$
C_{ij}
=
\begin{pmatrix}
624.707 & 23.729   & 325.332 & 8.34963 & 157.386 & 3.57778 \\
23.729  & 5.60873  & 11.6429 & 2.33996 & 6.39263 & 0.968056 \\
325.332 & 11.6429  & 905.777 & 29.3392 & 515.271 & 14.1013 \\
8.34963 & 2.33996  & 29.3392 & 5.42327 & 16.1422 & 2.85334 \\
157.386 & 6.39263  & 515.271 & 16.1422 & 1375.12 & 40.4327 \\
3.57778 & 0.968056 & 14.1013 & 2.85334 & 40.4327 & 6.25936
\end{pmatrix}.
$}
\end{equation}

\begin{figure*}
    \centering
    \includegraphics[width=0.45\linewidth]{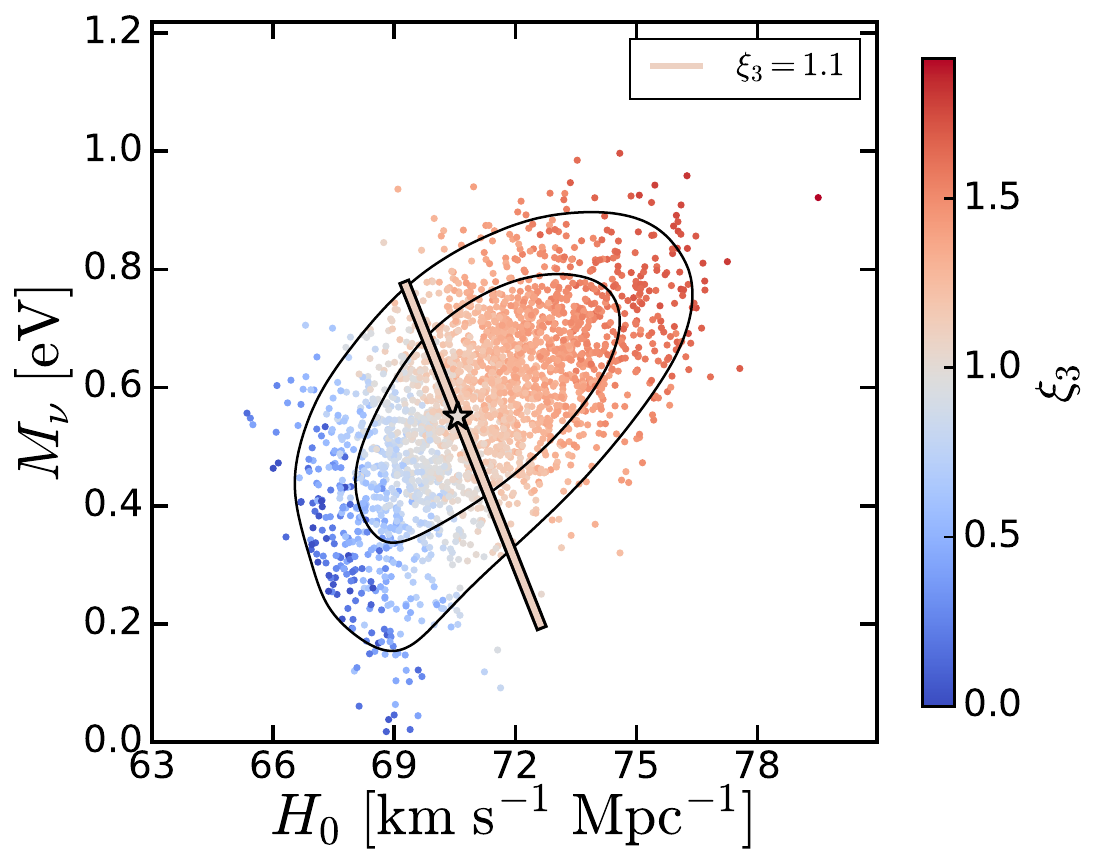}
    \includegraphics[width=0.45\linewidth]{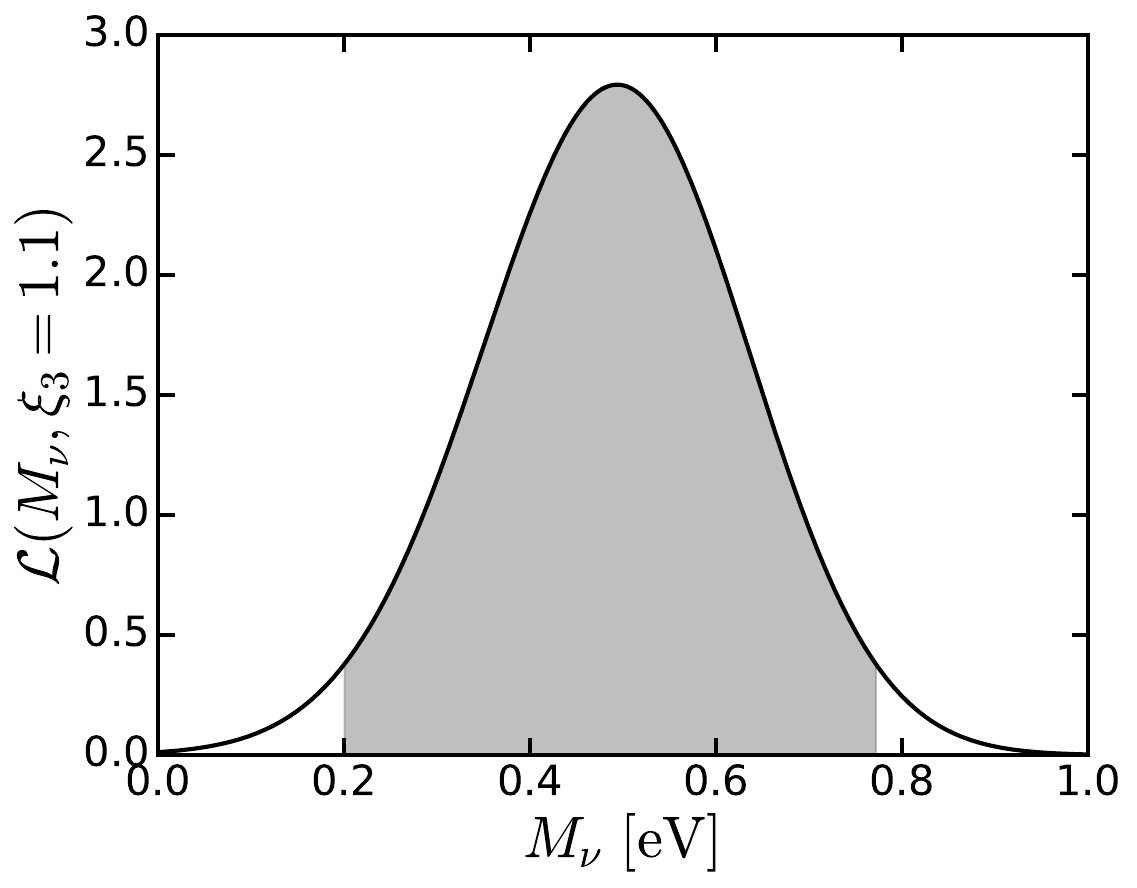}    
    \includegraphics[width=0.45\linewidth]{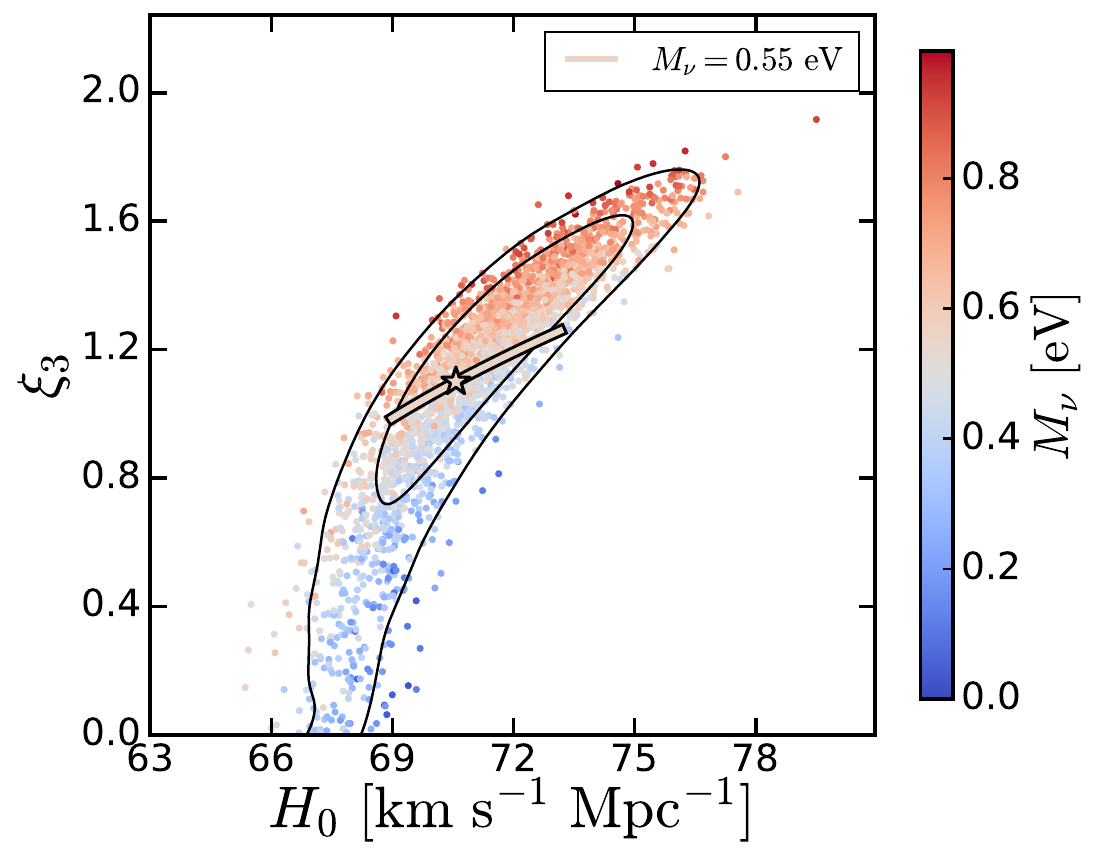}
    \includegraphics[width=0.45\linewidth]{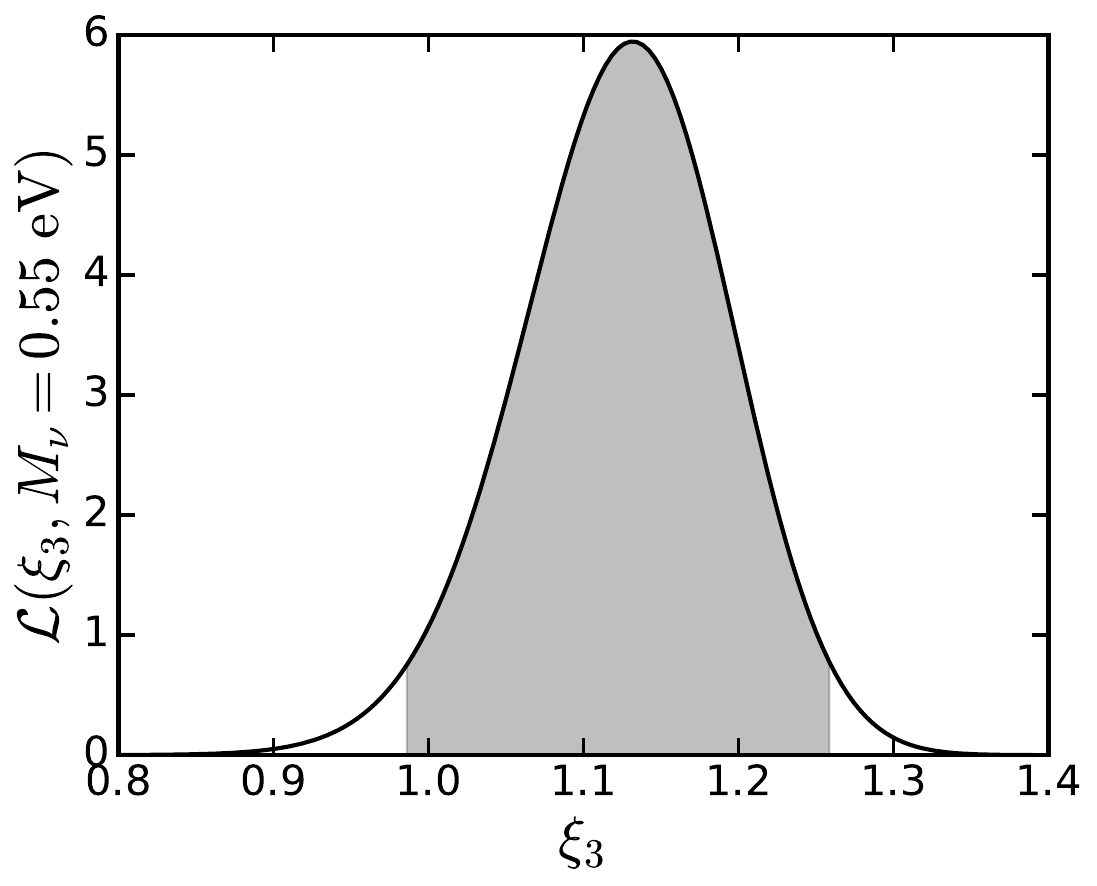}    
    \caption{Left panels are the same as the right panels of Figure~\ref{fig:taylor_results}, but with the light beige lines constrained by BAO data. The right panels show the corresponding normalized likelihood $\mathcal{L}$, with the shaded areas representing 95\% CL.}
    \label{fig:taylor_results_BAO}
\end{figure*}

Then, we can compute the likelihood function 
\begin{equation} 
    \mathcal{L}\sim e^{-\chi^2/2},
\end{equation} 
based on BAO observations and determine the 2$\sigma$ range of $\mathcal{L}$, as shown in the left two panels in Figure~\ref{fig:taylor_results_BAO}. The results for ($M_\nu$,$H_0$) and ($\xi_3$,$H_0$) are shown in the upper and lower right panels of Figure~\ref{fig:taylor_results_BAO}, respectively. This estimation aligns well with the 95\% contours in the MCMC results.



\end{document}